\documentclass[12pt]{article}
\usepackage{psfig}

\def\co{{\cal O}}
\def\vev#1{\left\langle #1 \right\rangle}
\def\del{{\partial}}
\def\href#1#2{#2}

\def\tf#1#2{{\textstyle{#1 \over #2}}}

\def\overleftrightarrow#1{\vbox{\ialign{##\crcr
     $\leftrightarrow$\crcr\noalign{\kern-0pt\nointerlineskip}
     $\hfil\displaystyle{#1}\hfil$\crcr}}}

\def\sqr#1#2{{\vcenter{\vbox{\hrule height.#2pt
         \hbox{\vrule width.#2pt height#1pt \kern#1pt
            \vrule width.#2pt}
         \hrule height.#2pt}}}}
\def\square{\mathop{\mathchoice\sqr56\sqr56\sqr{3.75}4\sqr34\,}\nolimits}

\begin{document}
\baselineskip=15.5pt
\pagestyle{plain}
\setcounter{page}{1}
\begin{titlepage}
\bigskip
\rightline{WIS/03/03-MAR-DPP}
\rightline{NSF-KITP-03-21}
\rightline{MIT-CTP-3352}
\rightline{UW/PT-03/06}
\rightline{hep-th/0303249}
\bigskip\bigskip\bigskip
\centerline{\Large \bf Defect Conformal Field Theory}
\medskip
\centerline{\Large \bf and Locally Localized Gravity}
\bigskip\bigskip

\centerline{\large Ofer Aharony$^a$, Oliver DeWolfe$^b$,}
\centerline{\large Daniel Z.\ Freedman$^{c}$ and
Andreas Karch$^{d}$}
\bigskip
\centerline{\em $^a$Department of Particle Physics, 
The Weizmann Institute of Science,}
\centerline{\em Rehovot 76100, Israel}
\centerline{\tt Ofer.Aharony@weizmann.ac.il}
\smallskip
\centerline{\em $^b$Kavli Institute for Theoretical Physics, UCSB, Santa
Barbara, CA 93106, USA}
\centerline{\tt odewolfe@itp.ucsb.edu}
\smallskip
\centerline{\em $^c$ Department of Mathematics and Center for
Theoretical Physics,} \centerline{\em Massachusetts Institute of
Technology, Cambridge, MA 02139, USA}
\centerline{\tt dzf@math.mit.edu}
\smallskip
\centerline{\em $^d$Department of Physics, University of Washington, 
Seattle, WA 98195, USA}
\centerline{\tt karch@feynman.phys.washington.edu}
\medskip
\bigskip


\begin{abstract}

Gravity may be ``locally localized'' over a wide range of length
scales on a $d$-dimensional anti-de Sitter ($AdS$) brane living inside
$AdS_{d+1}$.  In this paper we examine this phenomenon from the point
of view of the holographic dual ``defect conformal field theory''.
The mode expansion of bulk fields on the gravity side is shown to be
precisely dual to the ``boundary operator product expansion'' of
operators as they approach the defect.  From the field theory point of
view, the condition for localization is that a ``reduced operator''
appearing in this expansion acquires negative anomalous dimension.  In
particular, a very light localized graviton exists when a mode arising
from the reduction of the ambient stress-energy tensor to the defect
has conformal dimension $\Delta \sim d-1$.  The part of the stress
tensor containing the defect dynamics has dimension $\Delta = d-1$ in
the free theory, but we argue that it acquires a positive anomalous
dimension in the interacting theory, and does not participate in
localization in the regime of small backreaction of the brane.  We
demonstrate that such an anomalous dimension is consistent with the
conservation of the full stress-energy tensor.  Finally, we analyze
how to compute the anomalous dimensions of reduced operators from
gravity at leading order in the interactions with the brane.

\noindent
\end{abstract}
\end{titlepage}


\section{Introduction and Summary}

It was suggested a long time ago \cite{Rubakov} that the
$3+1$-dimensional world that we perceive around us could be merely a
subspace of a higher-dimensional non-compact universe, provided the
phenomena we observe are somehow localized to this subspace. The main
obstacle to realizing this suggestion lies in reconciling it with the nature
of gravity, which on the one hand is seen to display four-dimensional
behavior over the entire range of known experiments, but on the other
hand must propagate in all spacetime dimensions due to its geometrical
character.

In recent years it has been realized that this obstacle could be
overcome if the higher dimensional space is (asymptotically) anti-de
Sitter ($AdS$) space.  Randall and Sundrum demonstrated \cite{rs2}
that an effective lower-dimensional theory of gravity could be
localized on a flat brane in anti-de Sitter space, providing an
alternative to compactification.  This solution involves a fine tuning
between the brane tension and the bulk cosmological constant, so it
was natural to ask what happens to localized gravity as the tension is
detuned \cite{kaloper,kimkim,nihei,dfgk}.  A larger tension leads to a
de Sitter brane and more strongly localized gravity, but decreasing
the tension produces an anti-de Sitter brane and the localized
graviton apparently vanishes from the spectrum.  However, it was shown
in \cite{kr} that for a slight detuning, a massive mode of the
graviton with a hierarchically small mass approximates
lower-dimensional gravity over a range of length scales; this
phenomenon was dubbed ``locally localized gravity''. So far a ``local
localization'' of gravity by this phenomenon has only been observed in
the context of low-energy effective field theory. It is interesting to
study whether this phenomenon can also occur within a consistent
theory of quantum gravity, such as string theory.

The $AdS$/CFT correspondence \cite{juan,magoo} relates gravitational
phenomena in $(d+1)$-dimensional anti-de Sitter space to dynamics in a
non-gravitational $d$-dimensional conformal field theory, which can be
visualized as living on the boundary of the anti-de Sitter space.  For
the flat brane (as in \cite{rs2}) and de Sitter brane cases, the
solution for the $d$-dimensional brane in $AdS_{d+1}$ excises the
boundary from the spacetime, so it is subtle to analyze these
solutions in the context of the $AdS$/CFT correspondence. On the other
hand, the solution for an anti-de Sitter brane preserves the boundary
of the $AdS$ space-time, intersecting it on a codimension one
subspace.\footnote{Note that the $AdS$ brane is not itself a boundary
of the space-time.}  Consequently, it was natural to conjecture that
this geometry has a holographic description consisting of a field
theory containing a defect or hypersurface, also codimension one, on
which additional dynamics are localized.  Because the anti-de Sitter
isometries of the spacetime are broken to the smaller anti-de Sitter
group of the brane, the dual field theory preserves scale invariance
as part of a reduced conformal group.

An explicit realization of this correspondence in string theory
involving a D3/D5-brane system in type IIB string theory
was suggested in \cite{kr2}.  In this
set-up a stack of $M$ D5-branes wraps $AdS_4 \times S^2$ inside the
geometry $AdS_5 \times S^5$ with $N$ units of five-form flux. This is
the near-horizon limit of $N$ D3-branes intersecting orthogonally with
$M$ D5-branes along a $2+1$ dimensional subspace.  This system was
studied in detail in \cite{dfo}, where the Lagrangian for the dual
``defect conformal field theory'' (dCFT) was constructed, and
conformal invariance was proven for the theory with $U(1)$ gauge
group.  The superconformal symmetry for arbitrary gauge group was
proven by \cite{erdmenger1}.  Other studies of $AdS$/dCFT and defect
quantum field theories (dQFTs)
include \cite{sav2,bp,bbdo,Quella,Lee,Skenderis,Mateos,Yamaguchi,erdmenger2,
erdmenger3}; there are many additional papers on the special case of
$AdS_2$ branes in $AdS_3$.

One may then study whether the phenomenon of locally localized gravity
is realized in the D3/D5 system.  The gravitational approximation to
its dynamics (far away from the D5-branes) is valid when $N \gg 1$ and
$g_s N \gg 1$, where $g_s$ is the string coupling. Local localization
of gravity arises only when the backreaction of the branes,
proportional to $M$, is taken into account. However, thus far the
study of $AdS$/dCFT has only been possible in the limit of vanishing
backreaction, $M \ll N$, where the gravity background may be treated
as a set of probe D5-branes in a fixed anti-de Sitter space.  The
complete solution including the backreaction is unknown, although
steps toward understanding it have been taken in \cite{fayya}.  This
is an apparent obstacle to the search for local localization in string
theory.  The dual dCFT \cite{dfo}, on the other hand, is defined for
all $M$ and $N$, so one could hope to use the duality to study local
localization without finding the backreacted gravity solution.  As a
first step, one would like to address the question: what is the
realization of local localization in a dual field theory? In this
paper we will propose an answer to this question and justify our
proposal.

The essential point is the familiar correspondence between the mass of
a bulk mode and the conformal dimension of the dual operator.  The
$(d+1)$-dimensional graviton can be expanded in modes propagating on
the $d$-dimensional $AdS$ subspace, and local localization requires
that a mode should be present that has almost vanishing $AdS_d$ mass,
hence approximating the behavior of a $d$-dimensional graviton.  This
mode implies in the dual field theory the existence of a
$(d-1)$-dimensional spin-2 operator, resulting from some kind of
decomposition of the $d$-dimensional stress tensor, with conformal
dimension (under the reduced conformal group) approaching $\Delta =
d-1$.

From the field theory point of view there seem to be two natural
candidates for what this spin-2 operator, corresponding to the
``locally localized stress tensor'', should be.  Generally the full
stress tensor of a dQFT can be written in the form
\begin{eqnarray}
\label{StressTensor}
T_{\mu\nu}(\vec{y},x) = {\cal T}_{\mu\nu}(\vec{y},x) + \delta(x) \, 
\delta_k^\mu \delta_l^\nu
\, t_{kl}(\vec{y}),
\end{eqnarray}
where ${\cal T}_{\mu \nu}$ is the stress tensor in the absence of the
defect, composed of $d$-dimensional fields only, while $t_{kl}$ is the
contribution from defect interactions, including both $d$- and
$(d-1)$-dimensional fields.  (Here and below $\mu,\nu,\ldots$ run over
the $d$ coordinates of the ambient theory while $k,l,\ldots$ run over
the $(d-1)$ coordinates of the defect; we will label the defect
coordinates by $\vec{y}$ and the coordinate perpendicular to the
defect by $x$, with the defect at $x=0$.)  One candidate for the
locally localized stress tensor is a $(d-1)$-dimensional operator
arising in a power-series expansion of the stress tensor ${\cal
T}_{\mu \nu}$ of the $d$-dimensional fields around the defect.
Another natural candidate is simply $t_{kl}$.  We shall explore both
possibilities.\footnote{In the formula (\ref{StressTensor}) we assumed
that we have interactions localized on the defect; one could also
study theories where the defects just give ``boundary conditions'' for
the $d$-dimensional fields, as in the D3/NS5 system \cite{HW}, and
then the second term in (\ref{StressTensor}) will not appear and we
will only have the first possibility. We also assumed that we have the
same ambient theory on both sides of the defect, though this does not
have to be the case. All of our analysis in this paper will be valid
in these other cases as well, with obvious small modifications.}

The possibility of the localized stress tensor arising in an expansion
of ${\cal T}_{\mu\nu}$ is reminiscent of the localized graviton
appearing in the mode expansion of the full graviton in the original
model of \cite{kr}.  We will show that the two are indeed related.
There is a natural expansion for $d$-dimensional operators as they
approach the defect, namely the so-called boundary operator product
expansion (BOPE) that is used in quantum theories with planar
boundary.\footnote{The term ``defect operator product expansion''
(dOPE) was considered and prudently rejected.} In the absence of
interactions with the defect the BOPE is just the Taylor expansion of
an operator in the coordinate perpendicular to the defect, and the
expansion is just composed of the ``reduced operators'' $[\partial_x^n
\co](\vec{y}) \equiv \partial_x^n \co(\vec{y},x) |_{x=0}$.  The
interacting BOPE generalizes the Taylor series and introduces both
``reductions'' of other $d$-dimensional operators and
$(d-1)$-dimensional defect operators into the expansion.  We
demonstrate that the part of the BOPE involving reduced operators
matches precisely with the decomposition of $AdS_{d+1}$ bulk fields
into $AdS_d$ modes in the gravity theory.  The locally localized
graviton of the model of \cite{kr} can hence be identified with an
operator in the BOPE of the dual stress tensor; it is reasonable to
expect that the generic dual to a theory of ``locally localized gravity''
will evince similar behavior. For weak coupling with the defect, the
lowest dimension operator appearing in this BOPE has a dimension close
to $d$, while a very light graviton requires a dimension close to
$(d-1)$.  Thus, strong coupling with the defect is required to realize
locally localized gravity in this way, mirroring the requirement of
large backreaction in the gravitational theory.

We also consider the operator $t_{kl}$. When the field theory is free,
this operator has precisely the right dimension to correspond to an
{\em exactly} massless localized graviton, so at first sight it seems
a more natural source for localized gravity. The string theory
translation of $t_{kl}$ is different from a ``locally localized
graviton'', however -- instead, it is a spin-2 field living on the brane
itself. Such a field is ``naturally localized'' on the brane, in
contrast with the ``locally localized'' examples discussed above.  Even
though this possibility was not considered in \cite{Rubakov}, it has a
similar phenomenology to the locally localized case, and it seems
natural given our modern perspective of the many fields living
on branes, such as open string modes on D-branes. 

Naively, the conservation of the total stress tensor in
(\ref{StressTensor}) means that it has dimension $d$, and suggests
that $t_{kl}$ should have dimension $(d-1)$, meaning it should always
correspond to a localized massless graviton. However, we find that
despite the conservation of the total stress tensor, $t_{kl}$ alone is
not conserved and accordingly can acquire an anomalous dimension.  We
demonstrate this explicitly in a toy defect quantum field theory, as
well as showing how the related divergence is canceled in correlators
of the total stress tensor, as it must be since the total stress
tensor is conserved.

In the case of the D3/D5 system, in the regime where gravity is a good
description but the branes have small backreaction, there is no light
spin-2 field living on the D5-branes. Thus, we argue that in this
system $t_{kl}$ acquires a very large anomalous dimension at large
$g_s N$ and small $g_s M$, and hence does not lead to a light
localized graviton. This will always be the case in string theory in
the limit of small backreaction, since none of the branes of string
theory carry any light spin-2 fields in their worldvolume. Of course,
to achieve localization we necessarily need to have large backreaction
of the brane. In this regime it is possible that the field
corresponding to $t_{kl}$ could become light and become relevant for
localized gravity. In general this field mixes with the operators
coming from the expansion of ${\cal T}_{\mu\nu}$ near the brane, so
the localized graviton could also come from a mixture of the two types
of fields.

The next three sections of this paper will be devoted to justifying
the statements made above, and computing in field theory the anomalous
dimension of operators analogous to $t_{kl}$.  In the last section we
consider calculations in the supergravity limit that are useful for
computing the anomalous dimensions of reduced operators at large $g_s
N$ when the backreaction is small.  These involve the evaluation of
two-point functions for $d$-dimensional operators in the presence of
the brane. At leading order in the interactions with the brane, the
Feynman diagrams with sources at the boundary of $AdS$ space (``Witten
diagrams'') \cite{witten} generically have two internal integrations
to be performed, and consequently they are similar to four-point
functions in ordinary $AdS$/CFT.  We employ the techniques of
``without really trying'' \cite{wrt}, developed for calculating
four-point amplitudes, to evaluate the three types of diagrams
contributing to two-point functions at this leading order, for the
case of scalar fields.  All diagrams are seen to produce contributions
to the correlation functions with the form dictated by the reduced
conformal invariance.  In future work we hope to use these results to
study whether localization of gravity occurs in the D3/D5 system.

Anomalous dimensions appear in perturbation theory multiplying
logarithms of the coordinates in correlation functions.  One can
extract from four-point functions in a non-defect CFT the anomalous
dimensions of operators by analyzing the limit as the operators approach
each other in pairs using the ordinary OPE. Similarly, one may
extract the anomalous dimensions in our case by examining a two-point
function of $d$-dimensional operators in the limit as
the two operators approach the defect at fixed
distance from each other, using the BOPE. We find that
generically reduced operators obtain anomalous dimensions, which are
not related to the anomalous dimensions of the $d$-dimensional
operators that they reduce from. This is due to the fact that, in the
presence of the defect, the process of bringing an operator near the
defect leads to divergences and a regularization is required, as for
composite operators.

The organization of this paper is the following.  In section
\ref{bdod} we review and elaborate on some basic facts about dCFTs,
the BOPE, and conserved currents.  In section~\ref{mode_decomp} we
demonstrate that the decomposition of bulk $AdS_{d+1}$ fields into
$AdS_d$ modes directly implies the BOPE via the $AdS$/CFT correspondence. 
This leads us
to our proposal for the field theory interpretation of local
localization. In section~\ref{anom_dim} we perform field theory
calculations in a toy dQFT to demonstrate how the delta-function
supported part of a current can acquire an anomalous dimension without
violating conservation for the total current. Finally, in
section~\ref{ads_section} we turn to the gravity calculations
necessary to extract (in the appropriate limit) the anomalous
dimensions of reduced operators.  We evaluate the three possible
classes of two-point functions and interpret the logarithms that
appear in terms of anomalous dimensions.  An appendix summarizes some
conventions which are useful for section~\ref{anom_dim}.

\section{Ambient Operators and Defect Operators in dCFT}
\label{bdod}

\subsection{Generalities}

A $d$-dimensional conformal field theory has the spacetime symmetry
group $SO(d,2)$, which is also the isometry group of the $AdS_{d+1}$
background of its gravity dual.  Introducing a brane wrapped on an
$AdS_d$ submanifold, as in the system of \cite{kr}, breaks the
isometry group to $SO(d-1,2)$.  The $AdS_d$ brane intersects the
boundary in a codimension one hypersurface, inducing a spatial defect
in the field theory dual.  The symmetry group of the field theory is
thus broken to the $(d-1)$-dimensional conformal group; although
certain translations, rotations and special conformal transformations
are lost, scale invariance is preserved.  The dual is hence a defect
conformal field theory (dCFT). Of course, even though the study of
dCFTs is motivated by the fact that these theories naturally appear in
the context of the $AdS$/CFT correspondence, one can also in principle
construct such field theories directly.

It is important to note that the preserved symmetry group does not
only act on the directions along the defect; the scale transformation,
for example, rescales the perpendicular direction as well as the
defect directions, and the unbroken special conformal transformations
also involve the perpendicular direction.  The unbroken symmetries
constitute the subgroup of $d$-dimensional conformal transformations
preserving the defect $x=0$, which is the smallest group containing
the $ISO(d-2,1)$ Poincar\' e algebra as well as the $d$- (not
$(d-1)$-) dimensional inversion transformation $\{ x \to x / (x^2 +
\vec{y}^2); \vec{y} \to \vec{y} / (x^2 + \vec{y}^2) \}$.  Note that the
inversion must be about a point inside the defect.

In the case of supersymmetric theories, the brane/defect generically
breaks fermionic and R-symmetries as well.  Our canonical example will
be the defect superconformal theory descending from ${\cal N}=4$ super
Yang-Mills in four dimensions, which is dual to type IIB string theory
on $AdS_5\times S^5$ with $N$ units of five-form flux and $M$
D5-branes wrapped on $AdS_4\times S^2$ \cite{kr,dfo,HW}. In this case the
$SU(2,2|4)$ supergroup is broken to $OSp(4|4)$; besides the reduction
of the conformal/$AdS$ group from $SO(4,2)$ to $SO(3,2)$, the
R-symmetry is broken from $SU(4)$ to $SU(2)_V \times SU(2)_H$, and the
number of both ordinary and conformal supercharges is cut in half.

When a dCFT has a Lagrangian description, as is the case for this
example, the action can be written explicitly in terms of ``ambient''
fields $\phi_d(\vec{y},x)$ propagating throughout space as well as
``defect'' fields $\chi_{d-1}(\vec{y})$ confined to the defect.  The
ambient and defect fields are coupled on the defect:
\begin{eqnarray}
{\cal L} = {\cal L}_{amb}[\phi_d(\vec{y},x)] + \delta(x) \, {\cal L}_{def}
[\phi_d(\vec{y},0), \chi_{d-1}(\vec{y})] \,,
\end{eqnarray}
where ${\cal L}_{amb}$ is the Lagrangian of the parent theory in the
absence of the defect, and ${\cal L}_{def}$ contains couplings that
explicitly break the symmetries down to the preserved
subgroup.\footnote{There are also examples of dCFTs with codimension
bigger than 1, see for instance \cite{erdmenger2}. We will only
discuss the codimension 1 case here, though the generalization of our
results to arbitrary codimension should be straightforward.}

The defect fields transform under the preserved symmetry group as in
an ordinary conformal field theory.  From them we may form
gauge-invariant operators ${\cal O}_{d-1}(\vec{y})$ which will form
representations of the $(d-1)$-dimensional conformal algebra; these
operators could also contain the restrictions $\phi_d(\vec{y},0)$ of
ambient fields to the defect.  We can also assemble gauge-invariant
operators ${\cal O}_d(\vec{y},x)$ from ambient fields only, which are
defined throughout space.  The restrictions of these ambient operators
to the defect produces ``reduced operators'' $[\co_d](\vec{y}) \equiv
\co_d(\vec{y},x)|_{x \rightarrow 0}$.
In the free theory the
two types of $(d-1)$-dimensional operators, $\co_{d-1}(\vec{y})$ and
$[\co_d](\vec{y})$, are clearly
distinguishable, but once we add interactions they can mix together,
and we will call all of them ``defect operators'' (since they are
parameterized by a position along the defect). The correlation
functions of these operators among themselves obey the usual
constraints of $(d-1)$-dimensional conformal symmetry; in particular,
the two-point and three-point functions of primary operators are
determined (up to a constant) in terms of the scaling dimensions of
the operators, as in an ordinary CFT.

The constraints placed on correlation functions involving the ambient
operators $\co_d(\vec{y},x)$ by the defect conformal group were
analyzed in \cite{mcao} in the context of a conformal field theory
with a boundary, and the same analysis applies to our case.  The
reduced conformal algebra allows a non-zero one-point function for
primary scalar ambient operators ${\cal O}_d$ of scaling dimension
$\Delta$, of the form
\begin{eqnarray}
\label{onepoint}
\vev{\co_d(\vec{y},x)} = {A_{\co_d}
\over x^\Delta},
\end{eqnarray}
for some constant\footnote{In general we could have a different
constant appearing here for positive and negative values of $x$, and
we should write this as $A_{\co_d} / |x|^{\Delta}$ since $\Delta$ is
not necessarily integer. This is the form we would generally find from
explicit computations like the ones we perform in the later
sections. In the D3/D5 system which we will mainly be interested in,
the chiral operators in the bulk all have integer dimensions, and
there is a parity symmetry (discussed in \cite{dfo}) which relates
positive and negative values of $x$ and which allows only
even-dimensional operators to obtain one-point functions, so the
expression (\ref{onepoint}) is always valid and the one-point function is
always an analytic function of $x$. In
this particular system it seems that also the two-point functions
discussed below are always functions of $x$ (as we will write them)
and not of $|x|$. However, this need not be the case in general (and
it cannot be the case when non-integer dimensions are involved). We
will generally ignore this subtlety in this paper.} $A_{{\cal O}_d}$.
Mixed two-point functions of ambient operators of dimension $\Delta$
and defect operators of dimension $\Delta'$ are also uniquely
determined up to a constant, taking
the form
\begin{eqnarray}
\vev{\co_d(\vec{y},x) \, \co_{d-1}(\vec{y}')} = {B_{\co_d \co_{d-1}} \over
{x^{\Delta-\Delta'} (x^2 + (\vec{y}-\vec{y}')^2)^{\Delta'}}} \,,
\end{eqnarray}
for some constant $B_{\co_d \co_{d-1}}$.
The simplest allowed correlators of ambient fields which have some freedom
in their coordinate dependence are two-point functions of ambient scalar
operators, which have the form
\begin{eqnarray}
\label{ofourofour}
\vev{\co_d^1(\vec{y},x) \, \co_d^2(\vec{y}',x')} = {1  \over
{x^{\Delta_1} (x')^{\Delta_2}}} \, f(\xi) \,,
\end{eqnarray}
where $f(\xi)$ is an arbitrary function of the conformally invariant
variable
\begin{eqnarray}
\xi \equiv {{(\vec{y}-\vec{y}')^2 + (x-x')^2 }\over {4 x x'}}\,.
\end{eqnarray}

\subsection{Boundary operator product expansion}
\label{BOPESec}

Before we inserted the defect, the ambient operators ${\cal
O}_d(\vec{y},x)$ transformed in representations of the full
(super)conformal algebra. In the presence of the defect we can
decompose these into representations of the defect (super)conformal
algebra and hence it is natural to look for an appropriate way to
write any ambient operator in terms of a set of defect operators.

Since translations in $x$ are no longer a symmetry, we cannot use them
to relate operators at different values of $x$. We can still use the
scaling transformations to relate operators at different non-zero
values of $x$, so these are all in the same representation of the
conformal group. However, the reduced operators $[\co_d](\vec{y})
\equiv {\cal O}_d(\vec{y},x)|_{x \rightarrow 0}$ are no longer related
by any symmetry to the operators ${\cal O}_d(\vec{y},x\neq 0)$. In
particular, the reduced operators can mix with defect operators ${\cal
O}_{d-1}(\vec{y})$. It seems that even though the operators ${\cal
O}_d(\vec{y},x\neq 0)$ can have well-defined scaling dimensions, they
are never primaries of the defect conformal algebra, since they are
not annihilated by the preserved special conformal generators. Thus,
if we want to discuss the ${\cal O}_d$ using the usual representations
of the unbroken conformal algebra which are built from (quasi-)primary
operators, we need to do something else.

In order to discuss the behavior of ambient operators near the defect,
it is convenient to introduce the notion of an expansion in a series
of defect operators, which we will call the BOPE since it has similar
properties to the boundary operator product expansion discussed in
\cite{mcao} (and in many other places for two dimensional conformal
field theories). By the usual arguments of local field theories, we
can expand ambient operators as $x \to 0$ as a power series in
operators $\co_{d-1}$ localized at $x=0$, with a form dictated by
conformal invariance to be:
\begin{eqnarray}
\label{bope}
\co_d(\vec{y},x) = \sum_n {B^{\co_d}_{\co^n} \over
x^{\Delta_d-\Delta_n}} \co_{d-1}^n(\vec{y}).
\end{eqnarray}
The operators $\co_{d-1}^n$ appearing in this expansion can either be
made from defect fields, or from reductions of ambient fields to the
defect, or from both. For instance, if there are no interactions with
the defect, (\ref{bope}) is simply the Taylor expansion of the ambient
operator, and the operators $\co_{d-1}^n(\vec{y})$ are all of the form
$\partial_x^n \co_d(\vec{y},x) |_{x=0}$, of dimension
$\Delta_n=\Delta_d+n$. Once we add interactions these Taylor modes can
generally have anomalous dimensions and mix with defect
operators. Note that in general a reduced operator like $[\partial_x^n
\co_d](\vec{y}) \equiv \partial_x^n \co_d(\vec{y},x)|_{x \rightarrow
0}$ cannot be defined without subtracting a divergence.  The operators
$\co_{d-1}^n(\vec{y})$ appearing in (\ref{bope})
are generally primaries (or descendants of
primaries) of the defect conformal algebra; hence the natural way to
get primaries from an ambient operator after the introduction of the
defect is to compute its boundary operator product expansion.

One should note that in the free theory, the operators $[\partial_x^n
{\cal O}_d](\vec{y})$ are not precisely the primaries, though there
is a primary associated to each. This is obvious because the two-point
functions of these operators do not vanish for different $n$'s, even
though the operators with different $n$ have different scaling
dimensions. The actual primaries are of the form
\begin{eqnarray}
 [\widetilde{\partial_x^n {\cal O}_d}] \equiv [\partial_x^n {\cal O}_d] +
c_2 \nabla^2 [\partial_x^{n-2} {\cal O}_d] + c_4 (\nabla^2)^2
[\partial_x^{n-4} {\cal O}_d] + \ldots,
\end{eqnarray} 
where $\nabla^2$ is the $\vec{y}$-Laplacian.  Hence the BOPE expresses
${\cal O}_d$ in terms of both primaries and descendants of primaries.
The constants $c_i$, which depend on the dimension $\Delta$ of ${\cal
O}_d$, may be determined either by explicitly demanding annihilation
by the special conformal generators, or by requiring orthogonality of
the two-point functions. We will write some of them down in
section~\ref{nobrane}.

As discussed in \cite{mcao}, we can insert the BOPE into equations
such as (\ref{ofourofour}), and use the results to obtain information
about the dimensions of the operators living on the defect. The
two-point function (\ref{ofourofour}) has two different possible
limits -- small $\xi$ and large $\xi$. The small-$\xi$ limit
corresponds to bringing the operators close together away from the
defect, and it is natural to analyze this limit using the regular OPE
of the ambient theory; the behavior of $f(\xi)$ in this limit is
determined by the ambient OPE coefficients and by the one-point
functions (\ref{onepoint}) of the ambient operators $\co_d^3$
appearing in the OPE of ${\cal O}_d^1$ and ${\cal O}_d^2$. On the
other hand, the large-$\xi$ limit corresponds to bringing the
operators very close to the defect relative to their distance from one
another.  To analyze the behavior of $f(\xi)$ in this limit we plug
the BOPE (\ref{bope}) into equation (\ref{ofourofour}), for both
operators, leading to
\begin{eqnarray}
\label{largexi}
f(\xi) \simeq \sum_{n_1,n_2} 
B_{{\cal O}^{n_1}}^{\co_d^1} x^{\Delta_{n_1}}
B_{{\cal O}^{n_2}}^{\co_d^2} (x')^{\Delta_{n_2}}
\vev{\co_{d-1}^{n_1}(\vec{y}) \, \co_{d-1}^{n_2}(\vec{y}')} \,.
\end{eqnarray}
In the limit where the operators are very close to the defect, we have
$\xi \simeq (\vec{y}-\vec{y}')^2 / (4 x x')$. We see that the
contribution from a primary operator $\co_{d-1}^n$ appearing in the
BOPE\footnote{Normalizing the two-point function of $\co_{d-1}^n$ to be
$ \vev{\co_{d-1}^{n}(\vec{y}) \, \co_{d-1}^{n}(\vec{y}')} = 1 /
[(\vec{y}-\vec{y}')^2]^{\Delta_n}$.}  to the large-$\xi$ behavior of
$f(\xi)$ is $f(\xi) \simeq B_{{\cal O}^{n}}^{\co_d^1} B_{{\cal
O}^{n}}^{\co_d^2} (4/\xi)^{\Delta_n}$. Contributions from descendant
operators generally give higher powers of $1/\xi$ in the large-$\xi$
limit. Thus, from the large-$\xi$ behavior of (\ref{ofourofour}) we
can read off the dimensions (including the anomalous dimensions) of
all the defect operators $\co_{d-1}^n$ appearing in the BOPE of
both $\co_d^1$ and $\co_d^2$.

In the $AdS$/CFT correspondence, operators $\co_d(\vec{y},x)$ are mapped
to fields living in the full $AdS_{d+1}$ space, while operators
$\co_{d-1}( \vec{y})$ are mapped either to fields living on an $AdS_d$
brane or to the modes of the bulk fields (expanded near the brane).
We can compute all the correlators described above also on this
side of the correspondence, using string theory or using a
supergravity approximation including a brane source (which is valid
for large $g_s N$ and small $g_s M$). We will perform some
computations of this type in section \ref{ads_section}.

\subsection{Conserved currents}

Symmetries in a dCFT come in two types.  The simpler kind acts only on
the defect variables.  The associated currents are purely defect
operators, and the holographic duals to these currents are gauge
fields on the $AdS_d$ brane.  In the example of the D3/D5 system with
$M$ D5-branes, we have a global $U(M)$ symmetry acting on the $M$
hypermultiplets living on the defect, and the dual vector fields are
the zero modes of the gauge fields living on the D5-branes.  These
currents are conserved in the usual way, and do not acquire anomalous
dimensions.

The more interesting type of current corresponds to a symmetry acting
on both ambient and defect fields.  An example is the stress tensor,
or the R-currents in the supersymmetric systems.  Currents of this
type may be written in the form
\begin{eqnarray}
\label{tdecomp}
T_{\mu\nu}(\vec{y},x) &=& {\cal T}_{\mu\nu}(\vec{y},x) + 
\delta(x) \, \delta^k_{\mu} \delta^l_\nu \, t_{kl}(\vec{y}) \,, \\
J_{\mu}(\vec{y},x) &=& {\cal J}_{\mu}(\vec{y},x) + 
\delta(x) \, \delta^k_{\mu} \, j_k(\vec{y}) \,.\
\label{jdecomp}
\end{eqnarray}
These currents are dual to fields moving in the $AdS_{d+1}$ bulk, such as
the graviton or graviphotons.  The operators ${\cal J}$ and $j$ (or
${\cal T}$ and $t$) are not separately conserved, but the total $J$
obeys $\partial^{\mu} J_{\mu} = 0$ ($\partial^{\mu} T_{\mu \nu} = 0$)
both away from the defect and on the defect.
For the D3/D5 system, the fact that the energy-momentum tensor has the
form (\ref{tdecomp}) was noted in \cite{dfo}, and the $SU(2)_H\times
SU(2)_V$ R-currents similarly have the structure (\ref{jdecomp}).

In the free theory, ${\cal J}_\mu$ and $j_k$ are separately conserved;
$j_k$ has dimension $\Delta=d-2$ and ${\cal J}_\mu$ has dimension
$\Delta=d-1$. In the full theory, $j_k$ is no longer conserved, and
thus defect conformal invariance dictates that $j_k$ must have an
anomalous dimension, since a primary vector operator of $SO(d-1,2)$
can have dimension $\Delta=d-2$ if and only if it is a conserved
current.  A priori one might think that the appearance of such an
anomalous dimension in correlation functions like $\vev{J J}$ is
inconsistent with the conservation of $J$. However, we will show in
section~\ref{anom_dim} that the divergence in $\langle j \, j \rangle$
associated with the anomalous dimension is canceled in the full
$\langle J \, J \rangle$ correlator by contributions from $\langle
{\cal J} \, j \rangle$ and $\langle j \, {\cal J} \rangle$.  Similar
arguments apply to $T, {\cal T}$ and $t$, with the dimensions shifted
up by one.

In the next section, we will discuss obtaining a BOPE for operators
(including $T_{\mu\nu}$ and $J_\mu$) from gravity considerations.
Having done so, we will be in a position to put forward our proposal
for a definition of local localization in a dCFT.

\section{Local Localization and the BOPE}
\label{mode_decomp}

In the presence of an $AdS_d$ brane, it is natural to decompose
$AdS_{d+1}$ fields into modes transforming under the preserved $AdS_d$
isometry group.  This decomposition is analogous to the boundary
operator product expansion that we performed in the previous section for
field theory operators in the presence of the defect. In this section
we will show that the mode decomposition and the BOPE are actually
related by the $AdS$/CFT correspondence. This result will allow us to
provide a precise proposal for the realization of local localization
in dCFT.  For convenience, we perform the analysis of this section in
the Poincar\'e patch; the generalization to global $AdS$ is
straightforward.

\subsection{BOPE as gravity mode decomposition}
\label{BOPEgravity}

\begin{figure}
 \centerline{\psfig{figure=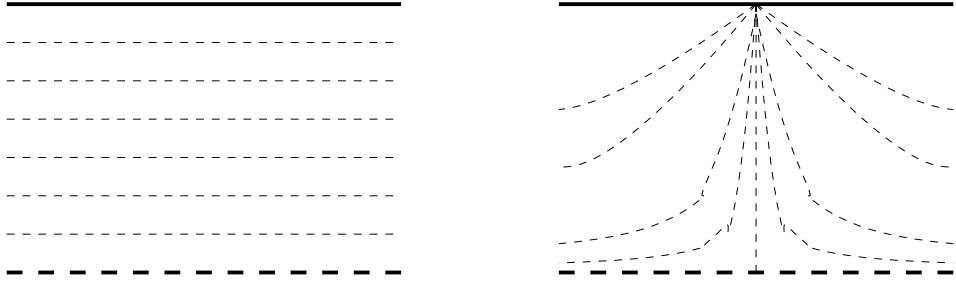,width=5in}}
 \caption{The two slicings of $AdS_5$. The horizontal axis is the
direction $x$ transverse to the brane and the vertical axis is the radial
direction of $AdS$ interpolating from the boundary (solid line) to the
horizon (dashed line).  The figure on the left shows lines of constant $\rho$
while the figure on the right shows lines of constant $r$.}
\label{ppatch}
  \end{figure}

In studying the $AdS$/CFT correspondence, it is standard to represent
$AdS_{d+1}$ as a foliation of $d$-dimensional Minkowski spaces. For
example, its metric may be written as
\begin{eqnarray}
\label{mink}
ds^2_{mink}
= L^2 \left ( e^{ 2 \rho} \left \{ 
d x^2 + d \vec{y}^2  \right \} +
d \rho^2 \right ) \,,
\end{eqnarray}
where $L$ is the curvature radius of $AdS_{d+1}$ and each
constant-$\rho$ slice is a copy of Mink$_{d}$.  On the other hand, in
analyzing the bulk physics of an embedded $AdS_d$-brane, it is useful
to express the geometry of $AdS_{d+1}$ in terms of a foliation of
$AdS_d$ submanifolds,
\begin{eqnarray}
\label{pads}
ds^2_{ads} = L^2 \left ( \cosh^2 r \left \{ e^{ 2 w}
d \vec{y}^2 + d w^2 \right \} + dr^2 \right ) \,, 
\end{eqnarray}
where each constant-$r$ slice is an $AdS_4$, with $-\infty < r < \infty$
and the brane at $r=0$.  The coordinate systems are related by
\begin{eqnarray}
\label{pchange}
x = e^{-w} \tanh r  \,, \quad \quad 
e^{\rho}=  e^w  \cosh r \,,
\end{eqnarray}
and are depicted in figure~\ref{ppatch}. From here on we will set
$L=1$ (it can always be restored by dimensional analysis).

Every $AdS_{d+1}$ bulk field $\phi_{d+1}(\vec{y},w,r)$ of mass $M$, 
transforming in some representation of $SO(d,2)$, decomposes into a tower of
$AdS_d$ modes $\phi_{d,n}(\vec{y},w)$ inhabiting representations of
the preserved isometry group $SO(d-1,2)$.  Each mode is multiplied by
an appropriate wavefunction of the $r$-direction:
\begin{eqnarray}
\label{Decompose}
\phi_{d+1}(\vec{y},w,r) = \sum_n \psi_n(r) \, \phi_{d,n}(\vec{y},w) \,.
\end{eqnarray}
Among the data of the $SO(d-1,2)$ representation is an $AdS_d$-mass
$m_n$ for each $\phi_{d,n}$,
\begin{eqnarray}
\partial_d^2 \phi_{d,n} = m_n^2 \phi_{d,n} \,,
\end{eqnarray}
where $\partial_d^2$ is the $AdS_d$-Laplacian.  The mass $m_n$ and the
wavefunction $\psi_n(r)$ may be determined by solving the wave
equation for $\phi_{d+1}(\vec{y},x,r)$.  In general the backreaction
of the brane may produce a more general warp factor $A(r)$, $ds^2 =
dr^2 + e^{2A(r)} ds_{AdS_d}^2$, although (\ref{pads}) will continue to
hold at large $|r|$; this more general metric still preserves $AdS_d$
isometries associated with dual dCFT.  To linear order the wave
equation then reduces to an ordinary differential equation for the
wavefunction $\psi_n(r)$,
\begin{eqnarray}
\label{wave}
\partial_r^2 \psi_n(r) + d A'(r) \partial_r \psi_n(r) + e^{-2A(r)} 
m^2_n \psi_n(r) - M^2 \psi_n(r)=0 \,.
\end{eqnarray}
This will receive corrections from various interactions in the brane
worldvolume theory,\footnote{The brane interactions will generally
cause a mixing between the modes corresponding to different bulk
fields $\phi_{d+1}$, though we neglect this here. However, precisely
the same phenomenon occurs also in the BOPE, and it is easy to
generalize our discussion to incorporate it.} all of which affect the
calculation of the masses $m_n$.

The field $\phi_{d+1}$ of mass $M$ is dual to an ambient operator
$\co_d(\vec{y},x)$ of dimension $\Delta_d$ (with $\Delta_d (\Delta_d -
d) = M^2$) in the dCFT.  Analogously, since the $\phi_{d,n}$ inhabit
an effective $AdS_d$ theory (they are representations of $SO(d-1,2)$),
they are related to dual ``defect operators'' $\co_{d-1,n}(\vec{y})$.
We wish to interpret the bulk relation (\ref{Decompose}) in terms of
some operator decomposition of $\co_d$ into the $\co_{d-1,n}$. In view
of the discussion in the previous section, it is natural to identify
the operators $\co_{d-1,n}(\vec{y})$ with the reduced operators
arising from the expansion of $\co_d$ near the defect.

In the $AdS$/CFT correspondence, the relation between bulk fields and
the values of operators in the ``boundary'' theory proceeds by
choosing a function $f$ on $AdS_{d+1}$ that has a pole on the
boundary, and multiplying each field by a power of this function
determined by the mass of the field to obtain finite boundary values
which are related to the dual operators \cite{witten}.  The freedom to
pick this function corresponds to the freedom to make conformal
transformations on the boundary theory; in particular the choice of
$f$ determines the metric on the boundary.

The expectation values of normalizable solutions for $\phi_{d+1}$ 
compute in this fashion vacuum
expectation values of the dual operators $\co_d$.
One possible choice of $f$, motivated by (\ref{pads}), is to pick 
the function $f$ to be the $AdS_d$-slicing
warp factor, $\cosh r$. For this choice we obtain the relation
\begin{eqnarray}
\label{dictr}
\phi_{d+1}(\vec{y},w,r) =
{(\cosh r)^{-\Delta_d} \over 2 \Delta_d - d}  \,  
\langle {\cal
O}_d(\vec{y},w) \rangle + \co((\cosh r)^{-(\Delta_d +2)}) \,,
\end{eqnarray}
near the boundary of $AdS_{d+1}$ at $r = \pm \infty$.  However, the
metric obtained on the boundary from this choice of $f$ will be
induced from the $AdS_d$ slicing, and hence will be two copies of
$AdS_d$, glued together along their common boundary, which is the
location of the defect.

This is not the boundary metric we desire, as we are interested in the
dCFT living on flat space (where our analysis of the previous section
applies).  The choice of function with pole $f$ that accomplishes this
is of course the warp factor for the Minkowski foliation, $e^\rho$.
This will induce a flat metric, and we shall see that it also induces
a nontrivial spatial variation transverse to the defect in the
boundary values of the wavefunctions $\psi_n(r)$.
In our choice of the flat boundary metric 
the relation between the bulk field and the expectation value of the
dual operator in the dCFT is hence
\begin{eqnarray}
\label{dict}
\phi_{d+1}(\vec{y},x,\rho) = {e^{-\Delta_d \rho} \over 2 \Delta_d - d}  \,  
\langle {\cal
O}_d(\vec{y},x) \rangle + \co(e^{-(\Delta_d +2)\rho}) \,,
\end{eqnarray}
as $\rho \rightarrow \infty$.  As usual, in the absence of a source
term for the field (which would be larger than the terms in
(\ref{dict}) for $\Delta_d \geq d/2$), the coefficient of the leading
term is identified with the VEV of the operator \cite{BKL}. We want to
read this as an operator equation, since it should still be true in
the presence of sources, as long as you are not on top of the source
-- in the presence of insertions of other operators one re-solves for
the perturbed gravity background, and again extracts the coefficient
of the same term as the value of the operator.

Now, $\phi_{d+1}$ also obeys the expression (\ref{Decompose}).
Consequently, using the coordinate relations (\ref{pchange}) evaluated
near the boundary, we may write
\begin{eqnarray}
\label{VevRelation}
\langle \co_d(\vec{y},x) \rangle = \lim_{\rho \rightarrow \infty} 
e^{\Delta_d \rho} (2 \Delta_d - d) \sum_n \psi_n(r) \, \phi_{d,n}(\vec{y},w) 
\bigg|_{ e^r=  2  x  e^\rho, \, e^{-w} = x}\,.
\end{eqnarray}
The wavefunctions $\psi_n(r)$ have a universal scaling near the
boundary
\begin{eqnarray}
\label{UniversalScaling}
\psi_n(r) = C_n (e^r)^{-\Delta_d} + \co(e^{-(\Delta_d + 2)r})\,,
\end{eqnarray}
regardless of $n$.  This is because for large $|r|$ the geometry
reverts to $AdS_{d+1}$ (\ref{pads}) and $e^{-2A(r)} \simeq \cosh^{-2}
r$ vanishes, meaning all modes are governed by
\begin{eqnarray}
\label{universal}
\partial_r^2 \psi_n + d A'(r) \partial_r \psi_n - M^2 \psi_n =0 \,,
\end{eqnarray}
leading to the result (\ref{UniversalScaling}).  The expression
(\ref{VevRelation}) then becomes
\begin{eqnarray}
\label{VevRelation2}
\langle \co_d(\vec{y},x) \rangle = (2x)^{-\Delta_d} (2 \Delta_d - d)
\sum_n C_n \, \phi_{d,n}(\vec{y},w) \bigg|_{e^{-w} = x}\,,
\end{eqnarray}
with some constants $C_n$ determined by the solution for $\psi_n(r)$.

The fields $\phi_{d,n}$ live in an $AdS_d$ space, and by the basic
principle of $AdS$/dCFT they are related to $(d-1)$-dimensional
operators $\co_{d-1,n}(\vec{y})$ living on the intersection of the
$AdS_d$ brane with the boundary.  The $\co_{d-1,n}$ have conformal
dimensions $\Delta_n$ determined by the $\phi_{d,n}$ masses as
$\Delta_n (\Delta_n - d + 1) = m_n^2$, and obey a relation completely
analogous to (\ref{dict}):
\begin{eqnarray}
\label{basic}
\phi_{d,n}(\vec{y},w) \simeq {e^{-\Delta_n w} \over 2 \Delta_n - d + 1} \,
\langle {\cal O}_{d-1,n}(\vec{y}) \rangle + \co(e^{-(\Delta_n +2)w})
\,
\end{eqnarray}
at large $w$.
However, there is an essential difference between the two equations for 
the purposes of our derivation.  While
the coordinate $e^\rho$ in (\ref{dict}) plays a purely holographic
role, and is eliminated in the process of computing boundary data, the
coordinate $e^w$ appearing in (\ref{basic}) is identified with $1/x$,
which survives in the field theory.  This is a consequence of our
order of limits, going to the boundary $\rho \rightarrow \infty$
before taking $x \rightarrow 0$.

As a result, the subleading terms in (\ref{basic}) all survive in the
boundary limit.  Using the $AdS_d$ wave equation
\begin{eqnarray}
(\partial_w^2 + (d-1) \partial_w + e^{-2w} \nabla^2 - m_n^2 ) 
\phi_{d,n}(\vec{y},w) = 0 \,,
\end{eqnarray}
where $\nabla^2$ is again the Laplacian in the $\vec{y}$ directions,
they may be calculated recursively in terms of
derivatives $\nabla^2$ of the VEV $\langle \co_{d-1,n}(\vec{y})\rangle$:
\begin{eqnarray}
\label{Recursive}
&& \phi_{d,n}(\vec{y},w) = e^{-\Delta_n w} \sum_{k=0}^\infty a_{k,n} \,
e^{-2kw} \, \nabla^{2k} \langle \co_{d-1,n}(\vec{y}) \rangle \,, \\
a_{k-1,n} &=& - \left[ ( \Delta_n + 2k)^2 - (\Delta_n + 2k)(d-1) - 
m_n^2 \right] a_{k,n} \,, \quad \quad  
a_{k=0,n} = {1 \over 2 \Delta_n - d+1} \,. \nonumber 
\end{eqnarray}
Hence we have
\begin{eqnarray}
\label{DeriveBOPE}
\langle \co_d(\vec{y},x) \rangle = (2x)^{-\Delta_d} (2 \Delta_d - d)
\sum_n C_n \, \sum_k   a_{k,n} \, x^{\Delta_n + 2k} \, \nabla^{2k} 
\langle \co_{d-1,n}(\vec{y}) \rangle \,.
\end{eqnarray}
Extrapolating this to an operator relation, we obtain precisely a
boundary operator product expansion of the form
\begin{eqnarray}
\label{bope2}
\co_d(\vec{y},x) = \sum_i {B^{\co_d}_{\co_i} \over
x^{\Delta_d-\Delta_i}} \co_{d-1,i}(\vec{y}) \,,
\end{eqnarray}
where $i$ runs over all values of $n$ and $k$, with the
$\co_i(\vec{y})$ corresponding to all $\nabla^{2k} \co_n(\vec{y})$, and
with $\Delta_i = \Delta_n + 2k$.

It is quite satisfying to derive the BOPE decomposition directly from
gravity considerations.  The analysis here was for scalar
operators/fields, but it can easily be generalized also to other
fields, and in particular to symmetric tensors which are relevant for
the localization of gravity.  We see directly that the BOPE of a
primary $d$-dimensional operator $\co_d(\vec{y},x)$ will generically
contain both $d-1$-dimensional primaries $\co_{d-1,n}(\vec{y})$ and
their descendants $\nabla^{2k} \co_{d-1,n}(\vec{y})$.  Moreover, the
relation (\ref{basic}) establishes the one-to-one correspondence
between modes $\phi_d$ in the $AdS$ decomposition (\ref{Decompose})
and the primaries $\co_{d-1,n}$ appearing in the BOPE.  Consequently,
we expect locally localized modes to be associated to particular
reduced primary operators appearing in the BOPE.

\subsection{Example of the no-brane case}
\label{nobrane}

It is illuminating to examine (\ref{DeriveBOPE}) in the elementary
case of a ``phantom'' brane.  This means that we study fields on
$AdS_{d+1}$ without including any interactions with a brane, but we
proceed with the decomposition into $AdS_d$ modes regardless.  As
discussed in section \ref{bdod}, for this trivial case the BOPE is
simply a Taylor expansion in $x$,
\begin{eqnarray}
{\cal O} (\vec{y}, x) = [{\cal O}] (\vec{y})
+ x [\partial_x {\cal O}] (\vec{y})
+ {x^2 \over 2!} [\partial_x^2 {\cal O}] (\vec{y}) + 
{x^3 \over 3!} [\partial_x^3 {\cal O}] (\vec{y}) + \ldots \,.
\end{eqnarray}
We need to bear in mind that the $[\partial_x^n \co]$ are
generically not primary operators for $n>1$, but rather each corresponds to the
sum of a primary and descendants.  For our explicit comparison, we
choose to study the $d=4$ case, for which the first few primaries
$\widetilde{[\partial_x^n \co]}$ at $n>1$ are
\begin{eqnarray}
\label{Primaries}
&&\widetilde{[\partial_x^2 {\cal O}]}\equiv[\partial_x^2 
{\cal O} + {1 \over 2 \Delta -1} \nabla^2 {\cal O}] \quad ,  
\quad 
\widetilde{[\partial_x^3 {\cal O}]}\equiv[\partial_x^3 
{\cal O} + {3 \over 2 \Delta +1} \nabla^2 \partial_x 
{\cal O}] \,, \\
&&\widetilde{[\partial_x^4 {\cal O}]}\equiv[\partial_x^4 
{\cal O} + {6 \over 2 \Delta +3} \nabla^2 \partial_x^2 
{\cal O} + {3 \over (2\Delta+1)(2\Delta+3)} \nabla^2 \nabla^2 
{\cal O}] \,, \nonumber
\end{eqnarray}
where $\co$ has conformal dimension $\Delta$.  In terms of these
primaries and their descendants we can rewrite the Taylor expansion
BOPE in the form
\begin{eqnarray}
\label{PrimaryBOPE}
{\cal O} (\vec{y}, x) &=& [{\cal O}] (\vec{y})
+ x [\partial_x {\cal O}] (\vec{y})
+ {x^2 \over 2} \widetilde{[\partial_x^2 {\cal O}] }(\vec{y})+ 
{x^3 \over 6} \widetilde{[\partial_x^3 {\cal O}]} (\vec{y})  - \\ && 
{x^2 \over 2(2 \Delta-1)} \nabla^2 [{\cal O}](\vec{y}) 
- {x^3 \over 2(2 \Delta +1)} \nabla^2 [\partial_x {\cal O}](\vec{y}) 
+ \ldots \,. \nonumber
\end{eqnarray}
We seek to derive this expansion from the gravity relation (\ref{DeriveBOPE}).

Let us now restrict ourselves further to the case of $d=\Delta=4$. For this 
case the wavefunctions for a ``phantom brane'' were
calculated in \cite{kr, por1}, and found to be
\begin{eqnarray}
\label{PhantomSolns}
\psi_n(r) &=& (\cosh^{-4} r) \;  _2F_1\left(\tf52 + \tf{n}{2}, -\tf{n}{2} ; 
3 ; \cosh^{-2} r \right) \\ &\sim& e^{-4\rho} {1 \over x^4} \,,
\nonumber
\end{eqnarray}
where $n=0,1,2 \ldots$ and $m_n^2 = (n+1)(n+4)$.
We consequently have $C_n=16$ in this case, and we
obtain from (\ref{VevRelation2})
\begin{eqnarray}
\label{VevRelationPhantom}
\langle \co(\vec{y},x) \rangle = 4 x^{-4} 
\sum_n \phi_{d=4,n}(\vec{y},w) \bigg|_{e^{-w} = x}\,,
\end{eqnarray}
where we set $\co_{d=4}(\vec{y},x) \equiv \co(\vec{y},x)$.
Let us first consider the primaries, which come from the leading term
in each $\phi_{d=4,n}$, as in (\ref{basic}).  The mass $m_n^2 =
(n+1)(n+4)$ implies $\Delta_n = 4 + n$, giving us from (\ref{basic})
\begin{eqnarray}
\phi_{d=4,n}(\vec{y},w) = {x^{4+n} \over 2n + 5 } \,
\langle {\cal O}_n(\vec{y}) \rangle + \ldots
\,,
\end{eqnarray}
with ${\cal O}_{d-1=3,n}(\vec{y}) \equiv {\cal O}_n(\vec{y})$.  Since
the $\co_n$ are primaries, they must correspond to the
$\widetilde{[\partial_x^n \co]}(\vec{y})$ up to a possible
normalization, $\co_n(\vec{y}) \equiv \beta_n \,
\widetilde{[\partial_x^n \co]}(\vec{y})$, leading to the primary part
of the BOPE
\begin{eqnarray}
\label{VevRelationPhantom2}
\co(\vec{y},x) =  
\sum_n {4 \beta_n \over 2n + 5}  \, x^n \, 
\widetilde{[\partial_x^n \co]}(\vec{y}) \, + {\rm descendants} \,.
\end{eqnarray}

For this to match (\ref{PrimaryBOPE}) we must have $4 \beta_n/ (2n+5)
= 1/n!$. In order to verify this we need to compare the normalization
of the $\co_n(\vec{y})$ to that of the $\widetilde{[\partial_x^n
\co]}(\vec{y})$.  A canonically normalized $AdS_d$ scalar with
dimension $\Delta$ gives rise to a dual operator with the two-point
function
\begin{eqnarray}
\label{CanonicalNorm}
\langle \co(\vec{y}) \, \co(0) \rangle = {1 \over
\pi^{(d-1)/2}} (2 \Delta - d+1) {\Gamma(\Delta) \over \Gamma(\Delta -
\tf{d-1}{2})} {1 \over y^{2\Delta}} \,.
\end{eqnarray}
However, the $\phi_{d,n}$ are in general not canonically normalized;
their normalization in the action is determined by dimensionally
reducing the $AdS_{d+1}$ action for $\phi_{d+1}$, and this leads to an
extra factor of $\int dr e^{2A(r)} \psi_n(r)^2$ outside the kinetic terms
for $\phi_{d,n}$, which then appears in the denominator of the
two-point function.  For the case of the $d=\Delta=4$ phantom brane, we can
compute this by integrating (\ref{PhantomSolns}) to obtain
\begin{eqnarray}
\langle \co_n(\vec{y}) \, \co_n(0) \rangle = 
{(2n+5) (n+4)! \over 128 \, n! } \times 
{1 \over
\pi^{3/2}} (2 n + 5) {\Gamma(n + 4) \over \Gamma(n +
\tf52)} {1 \over y^{8 + 2n}} \,.
\end{eqnarray}
Thus, the coefficients of the BOPE only come out right if it happens that
\begin{eqnarray}
\label{NormFormula}
\langle \widetilde{[\partial_x^n\co_n]}(\vec{y}) \, 
\widetilde{[\partial_x^n\co_n]}(0) \rangle &=& {1 \over \beta_n^2} 
\langle \co_n(\vec{y}) \, \co_n(0) \rangle \\
&=& {n! (n+3)! (n+4)! \over 8 \pi^{3/2} \Gamma(n + \tf52)} 
{1 \over y^{8 + 2n}} \,. \nonumber
\end{eqnarray}
We have not been able to compute the two-point functions of the
primaries for general $n$, but we have checked that the first 4
primaries listed in (\ref{Primaries}) do satisfy (\ref{NormFormula})
exactly, assuming that the 4D $\co(y,x)$ has the canonical two-point
function given as the $AdS_{d+1}$ version of (\ref{CanonicalNorm})
with $d=\Delta=4$.  Hence the Taylor series structure is exactly
reproduced for the primaries up to this order.  We may regard
(\ref{NormFormula}) as a prediction for the norms of the higher
primaries.

One is then left needing to match the descendants; the terms $a_{k,n}$
with $k \neq 0$ in the expansion (\ref{Recursive}) must produce the
second line of (\ref{PrimaryBOPE}).  We have verified that this is
indeed the case up to $\co(x^4)$.  The $\co(x^4)$ term is the most
intricate, as the Taylor term $[\partial_x^4 {\cal O}]$ is composed of
a combination of the primary $\widetilde{[\partial_x^4 {\cal O}]}$ as
well as the $k=1$ descendant of $\widetilde{[\partial_x^2 {\cal O}]}$
and the $k=2$ descendant of $[\co]$.  Higher terms will be more
complicated still.

To summarize this subsection, 
we have verified explicitly to fourth order in $x$ that
the mode decomposition in the presence of the phantom brane is
precisely equivalent to a simple Taylor series for the dual field
theory operator for the case $d=\Delta=4$.  We expect this agreement
to persist to all orders, and to pertain also to other cases.  In
particular, when there is a genuine interacting brane, we expect
(\ref{DeriveBOPE}) to produce the reduced operators in the full
interacting BOPE for the dual dCFT.

\subsection{Local localization in dCFT}

Local localization occurs when one of the modes in the decomposition
of the graviton, 
\begin{eqnarray}
h_{kl}(\vec{y},r,w) = \sum_n \psi_n(r) \, (h_{kl})_{d,n}(\vec{y},w)\,,
\end{eqnarray}
has an $AdS_d$ mass that is hierarchically small compared to the scale
of the $AdS$ curvature, which characterizes the masses of the other
fields.  For the simplest system of a gravitating 4D brane coupled to
5D gravity with a cosmological constant \cite{kr}, the existence of
the locally localized mode has been confirmed both numerically
\cite{kr, matthew} and analytically \cite{miemiec,por1}.  One may
expect that local localization appears in many other gravitating
systems as well.  One way to try to confirm this is to explicitly
construct such systems, in string theory for example; this has thus
far proven difficult (see \cite{fayya} for the state of the art in the
D3/D5 system).  Since every gravity phenomenon should have a field
theory interpretation, even novel ones like local localization,
another way to proceed is to abstract the concept to an effect in the
dual field theory, and then look for local localization in the dCFT
duals even when the exact gravity solution including brane
backreaction is not known.  Understanding the nature of local
localization in the dCFT dual is also interesting in its own right,
and this is what we attempt to do here.

The analysis of subsection~\ref{BOPEgravity} can be readily adapted to
the study of the graviton, as the transverse polarizations obey a
massless scalar wave equation.  It is easy to verify that the
wavefunctions discussed above are localized near the brane if and only
if $\Delta_n < \Delta_d$; in particular near the boundary the
wavefunctions behave as $x^{\Delta_n-\Delta_d}$.  Thus, we can say
that a mode $\phi_{d,n}$ begins to be localized around the brane
whenever its mass is such that the associated scaling dimension
$\Delta_n$ is smaller than the scaling dimension $\Delta_d$ associated
with the parent field $\phi_{d+1}$.

Through the equivalence of the $AdS$ mode expansion and the BOPE that
we have established, and by using the usual $AdS$/CFT relation between
mass and conformal dimension, we can naturally recast this result in a
field theory language as a relation between anomalous dimensions.  In
particular, when local localization of gravity occurs, a reduced
operator in the BOPE expansion of the $d$-dimensional stress tensor
has a conformal dimension approaching $d-1$. As the dimension becomes
closer to $d-1$, gravity will be localized to the brane over a larger
and larger range of length scales.  Note that a dimension $\Delta_n <
d-1$ is forbidden by unitarity, and $\Delta_n = d-1$ can occur only if
the reduced stress tensor is conserved on its own and the theory has
two decoupled sectors.

We can therefore make a natural definition :

{\em Local localization of gravity occurs in a defect conformal field
theory when there is an operator in the BOPE of the stress tensor with
$\Delta - (d-1)  \ll 1$.}

In more generality, one can abstract the notion to a generic localized
field in a dCFT; the corresponding operator should have in its BOPE a
defect operator with a conformal dimension smaller than that of the
parent $d$-dimensional operator.

Obviously, local localization does not occur when we have no
interactions with the defect.  This is evident since the BOPE in that
case is just the Taylor expansion, and the conformal dimensions of the
reduced operators are simply $\Delta_d + n$, $n = 0, 1, 2, \ldots$,
none of which is smaller than $\Delta_d$.  As the coupling is turned
on, localization will occur if the $n=0$ reduced operator acquires a
negative anomalous dimension.  This is another example of a familiar
property of $AdS$/CFT, that phenomena in classical gravity can be
mapped to quantum effects in the large-coupling limit of the dual
field theory.

Our discussion in this section so far was limited only to the
``reduced operators'' arising as modes of the ambient operators.  At
first sight, it seems that another possible candidate spin-2 operator
to have a scaling dimension close to $\Delta=d-1$ is the
defect-localized part of the stress tensor $t_{kl}$ (\ref{tdecomp}).
In the free theory, its dimension is exactly $\Delta = d-1$, and since
it is a part of a conserved current, one may be tempted to believe its
dimension is not renormalized and remains $d-1$.

This point of view is problematic, however.  The defect part of the
stress tensor is a fundamentally defect operator, not a moment of an
ambient one; consequently it is expected to be dual to dynamics
localized on the $AdS_d$ brane itself, rather than to bulk physics.
In the well-understood probe regime of the D3/D5 system, with $g_s N$
large and $g_s M$ small, $t_{kl}$ is therefore dual to a spin-2 open
string field localized on the D5-branes.  Since there are no massless
spin-2 open string modes, the dual of $t_{kl}$ must be a massive
stringy mode, and thus it should acquire a large anomalous dimension
$\Delta \simeq (g_s N)^{1/4}$.  On the other hand, the dimensions of
the BOPE operators coming from ${\cal T}_{\mu\nu}$ are still close to
$\Delta^{(n)} = 4+n$ in this regime, since the backreaction of the
brane is small.  Hence the ``naturally localized'' field $t_{kl}$ is
not a part of the onset of locally localized gravity in the small-$g_s
M$ regime.  Similarly, the defect parts $j_k$ of the $SU(2)_H\times
SU(2)_V$ R-symmetry currents (\ref{jdecomp}) should also acquire
anomalous dimensions in the D3/D5 system, as they sit in the same
multiplet as the energy-momentum tensor.

In more general systems, we expect the story to be similar.  There is
no example in string theory of a light spin-2 brane mode that could be
dual to $t_{kl}$, and hence we expect the dual field to generically be heavy
in regimes where a brane description is adequate.  In
section~\ref{anom_dim} we demonstrate that the defect component of a
current indeed acquires an anomalous dimension in a simple dQFT, and
we show how this is consistent with the conservation of the full
current. Similar computations could also be done for the stress
tensor.

In section~\ref{ads_section}, we perform the calculations necessary
for the supergravity computation of the anomalous dimensions of
reduced scalar operators, to leading order in $g_s M$.  We shall see
that anomalous dimensions generically appear here as well, but since
they are corrections of order $g_s M$, they are not large in the probe
limit. We expect that similar computations will give the anomalous
dimensions of the BOPE modes of $T$; in the D3/D5 system these
anomalous dimensions are related by supersymmetry to those of some reduced
scalar operators, so the computation of the latter suffices for the
analysis of locally localized gravity.  Such computations should
reveal whether there is an onset of localization as the backreaction
of the brane starts to be taken into account.

The hope for a truly localized graviton lies, of course, with strong
backreaction, that is large $g_s M$ as well as large $g_s N$. In this
regime we do not know how to compute the anomalous dimensions of the
various fields, and we do not know whether the dimension of $t$ is
smaller or larger than that of the BOPE modes coming from ${\cal
T}$. Since we show in this paper that both $t$ and the modes coming
from ${\cal T}$ have anomalous dimensions, it is possible that one of
these operators has a dimension such that $\Delta -(d-1) \ll 1$ at large
$g_s M$, so that localized gravity would be realized in string theory
in this regime. It is hard to say whether $t$ could come down in
dimension at large $g_s M$ and participate in the local localization,
or if this localization necessarily involves only a mode of ${\cal
T}$.  For large $g_s M$ there could in any case be a large mixing
between $t$ and the modes of ${\cal T}$, so the distinction may well
be meaningless.

\section{Anomalous Dimensions of Defect Operators}
\label{anom_dim}

In this section we wish to verify that the picture we have developed,
in which the $(d-1)$-dimensional component of the stress tensor
acquires an anomalous dimension despite the fact that the full stress
tensor containing it is conserved, is consistent. For simplicity,
rather than analyzing the stress tensor in the D3/D5 system, we pick a
toy model involving a subset of the fields of the D3/D5 system, and we
examine the same phenomenon for a conserved current in this toy
model. We expect the behavior of the stress tensor itself (which, in the
D3/D5 system, is related by supersymmetry to the currents) to be
analogous.

As our toy model, we consider an elementary
dQFT (not a conformal theory) given by the Lorentzian action
\begin{eqnarray}
\label{Action}
S_{dQFT} = \int d^4z & [ &
-\tf{i}{4} \bar\lambda^{mi} \gamma^\mu \partial_\mu \lambda^{im}
+ \delta(x) \left(-i \bar\Psi^i \rho^k \partial_k \Psi^i + 
\partial^k \bar{q}^m \partial_k q^m \right) \\
&+&  g \delta(x) \left(i \bar{q}^m (\bar{\lambda}_1)^{mi} \Psi^i - i
\bar{\Psi}^i (\lambda_1)^{im} q^m \right)] \,.
\nonumber
\end{eqnarray}
Here $q$ is a complex scalar, $\Psi$ is a complex 3D spinor and
$\lambda$ is a 4D Majorana spinor; $\lambda_1$ is the projection of
$\lambda$ onto two of its components, see \cite{dfo}. In this section
we use $z^\mu$ as a shorthand for $(y^k, x)$.  The theory
(\ref{Action}) is a subset of the dynamics of the D3/D5 dCFT
\cite{dfo}, keeping only the fermions in the ambient four-dimensional
theory.  The $SU(2)_V \times SU(2)_H$ currents associated to the
global symmetry (which is the R-symmetry in the full dCFT) are
\begin{eqnarray}
\label{currents}
J_V^{A \mu} &=& \tf14 \bar\lambda^{mi} \gamma^\mu (T^A)_{ij} \lambda^{jm} + 
\delta(x)\,  \delta^\mu_k \, \bar\Psi^i \rho^k (T^A)_{ij} \Psi^j \,, \\
J_H^{I \mu } &=& - \tf14 \bar\lambda^{mi} \gamma^\mu \lambda^{in} (T^I)_{nm}  
-i \delta(x) \,\delta^\mu_k \, \bar{q}^m 
\overleftrightarrow{\partial^k} (T^I)_{mn} q^m \,,
\nonumber
\end{eqnarray}
which are of the form (\ref{jdecomp}). The matrices $T^A$ and $T^I$
here are the generators of the global symmetry group; the rest of the
notations are explained in the appendix. The $SU(2)_H$ indices $m,n$
will play no role in our computation in this section, but we keep them
so that (\ref{Action}) is a subtheory of the D3/D5 dCFT.

We shall study the two-point function of $J_V^\mu$ to two-loop order,
and we drop the subscript $V$ henceforth.  We work directly in
coordinate space and resolve divergences using differential
regularization \cite{diffreg}.  We are interested in properties of the
correlator $\langle J(z_1) J(z_2) \rangle$ for separated points, so
only subdivergent parts of 2-loop diagrams are relevant.  We will
demonstrate that the defect current $j$ is not protected, and we
compute its anomalous dimension.  The correlator $\langle j(y_1) \,
j(y_2) \rangle$ thus depends on the renormalization scale $M$.  We
will show that the scale-dependence cancels in the full correlator
$\langle J(z_1) \, J(z_2) \rangle$, consistent with the conservation
of $J$.  This comes about because $\langle {\cal J} \, j \rangle$
contains a subdivergence localized on the defect, leading to a scale-dependent
contribution which provides the needed cancellation. To the order
considered here $\langle {\cal J} \, {\cal J} \rangle$ is UV finite and
does not participate in the cancellation.

To perform the computations, we find it convenient to employ a 4D
spinor notation.  A 4D spinor index decomposes into a 3D spinor index
$\alpha=1,2$ and an internal index $a=1,2$; only the $a=1$ component
of $\lambda$ participates in the defect interaction.  Define the 4D
spinor
\begin{eqnarray}
\Psi^i_{\beta b} \equiv \Psi^i_{\beta} \delta_{1b} \,.
\end{eqnarray}
The interaction in the Lagrangian (\ref{Action}) then becomes
\begin{eqnarray}
\label{4DInt}
S_{dQFT} &\supset& \int d^4z \, g \delta(x) \left(i \bar{q}^m 
\bar{\lambda}^{mi} \Psi^i - i
\bar{\Psi}^i \lambda^{im} q^m \right) \,,
\end{eqnarray}
where $\lambda$ and $\Psi$ are both 4D spinors.  The propagator of the
4D $\Psi$ then contains the projection matrix $P_+ \equiv (1 + \gamma^5
\gamma^x)/2$, as detailed in the appendix.

\begin{figure}
 \centerline{\psfig{figure=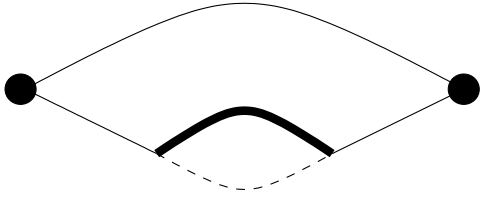,width=2.5in}}
 \caption{The leading diagram contributing to the $\langle j j \rangle $
 current-current correlator. }
\label{c1}
  \end{figure}

\begin{figure}
 \centerline{\psfig{figure=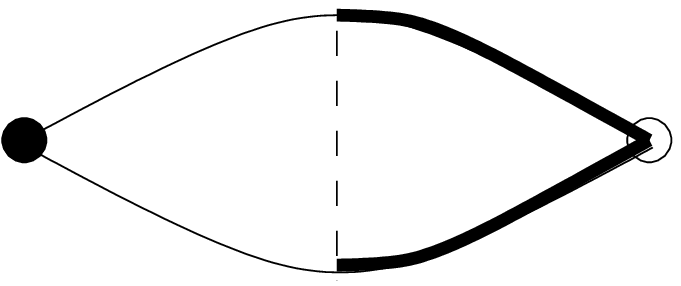,width=2.5in}}
 \caption{The leading diagram contributing to the $\langle {\cal J} j \rangle$
 current-current correlator. }
\label{c2}
  \end{figure}

\begin{figure}
 \centerline{\psfig{figure=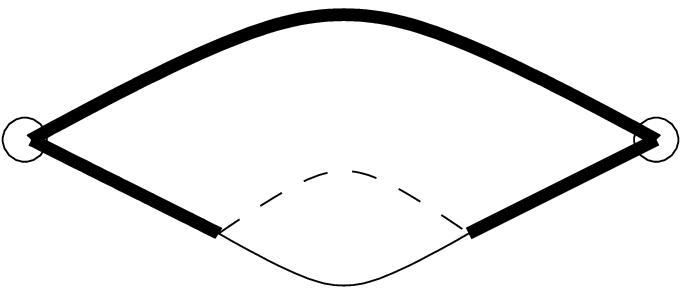,width=2.5in}}
 \caption{The leading diagram contributing to the $\langle {\cal J} {\cal J}
\rangle$ current-current correlator. }
\label{c3}
  \end{figure}

We study the order $g^2$ contributions to the correlator
\begin{eqnarray}
\label{Corr}
\langle J^{Ak}(z_1) \, J^{Bl}(z_2) \rangle &\equiv&
\langle {\cal J}^{Ak}(z_1) \, {\cal J}^{Bl}(z_2) \rangle + \delta(x_2)
\langle {\cal J}^{Ak}(z_1) \, j^{Bl}(y_2) \rangle \\ &+& \delta(x_1)
\langle j^{Ak}(y_1) \, {\cal J}^{Bl}(z_2) \rangle + \delta(x_1) \delta(x_2)
\langle j^{Ak}(y_1) \, j^{Bl}(y_2) \rangle \,.
\nonumber
\end{eqnarray}
Here, $A$ and $B$ are indices in the adjoint of $SU(2)_V$, and $k,l$
are $d=3$ vector indices. Figures \ref{c1}, \ref{c2} and \ref{c3} show
the lowest-order diagrams involving interactions that contribute to
correlators of $\vev{j j}$, $\vev{{\cal J} j}$ and $\vev{{\cal J}
{\cal J}}$, respectively.  In the figures we use dashed lines for
scalars and unbroken lines for fermions, thick lines for ambient
fields and thin lines for defect fields. The filled dots correspond
to insertions of $j$ and the open dots to insertions of ${\cal J}$. 
All results are stated for the Euclidean continuation of the 
correlation functions.

In the $\langle j \, j \rangle$ diagram of figure \ref{c1} 
we have a $\Psi$ loop where
one $\Psi$ propagator contains a $\lambda/q$ self-energy loop.
The result is
\begin{eqnarray}
\label{33}
\langle j^{Ak}(y_1) \, j^{Bl}(y_2) \rangle  =
 -8 \,g^2 \,{\rm Tr}\, (T^A T^B) 
\int d^3y_3 \, d^3y_4 \, \delta_{34} \, {\rm Tr}\, [ \gamma^k \, 
\hat{s}_{12} \, \gamma^l \, \hat{s}_{23} \, S_{34} \, \hat{s}_{41}] \,,
\end{eqnarray}
where $S_{34}$ denotes the fermion propagator from $z_3 = (\vec{y_3},x=0)$
to $z_4$, and similarly for the other propagators; see the appendix.
The coefficient comes from a $2$ for the two diagrams, a $2$ for the
$SU(2)_H$-index loop and a $2$ for the index reduction of the
$\lambda$ propagator.

In $\langle {\cal J} j \rangle$, we have an exchange diagram where a
$\Psi \Psi q$ triangle and a $\lambda \lambda q$ triangle share the
$q$ propagator, drawn in figure~\ref{c2}.  We find
\begin{eqnarray}
\label{43}
\langle {\cal J}^{Ak}(z_1) \, j^{Bl}(y_2) \rangle  =
 -4 \,g^2 \,{\rm Tr}\, (T^A T^B) 
\int d^3y_3 \, d^3y_4 \, \delta_{34} \, {\rm Tr}\, 
[ S_{31} \, \gamma^k \, S_{14} \, \hat{s}_{42} \, \gamma^l \, \hat{s}_{23}] \,.
\end{eqnarray}
$\langle j \, {\cal J} \rangle$ follows directly.  The numerical factor 
includes $1/4$ for the ${\cal J}$ current, $2$ for two different Majorana
contractions, $2^2$ for the $SU(2) \times SU(2)$ index reduction in
the two $\lambda$ propagators as in (\ref{FlavorProp}), and $2$ for the
$SU(2)_H$-index loop.

In $\langle {\cal J} {\cal J} \rangle$, the leading diagram is
the $\lambda$ loop with a self-energy interaction, drawn in figure~\ref{c3}.
Due to the Majorana condition there are 8 distinct Wick contractions, 
which in the end each contribute equally to the amplitude
\begin{eqnarray}
\label{44}
\langle {\cal J}^{Ak}(z_1) \, {\cal J}^{Bl}(z_2) \rangle  = 
-8  \,g^2 \,{\rm Tr}\, 
(T^A T^B) \int d^3y_3 \, d^3y_4 \, \delta_{34} \, {\rm Tr}\, [S_{41} \, 
\gamma^k \, S_{12} \, \gamma^l \, S_{23} \, \hat{s}_{34} ] \,.
\end{eqnarray}
The factor of $8$ can be understood as follows: $(1/4)^2$ for the
definition of the two currents, $8$ for the eight separate terms,
$2^3$ for the $SU(2) \times SU(2)$ index reduction in the three
$\lambda$ propagators as in (\ref{FlavorProp}), and $2$ for the
$SU(2)_H$-index loop.

We now proceed to the evaluation and regularization of these diagrams.
The first point to notice is that there is no subdivergence in
$\langle {\cal J} {\cal J} \rangle$, and the $y_3,y_4$ integrals
converge for separated external points.  

Next we analyze $\langle j j \rangle$. In detail the amplitude is
\begin{eqnarray}
\label{jjcorr}
\langle j^{Ak}(y_1) \, j^{Bl}(y_2) \rangle  =
 -{ 8 \,g^2 \over (4 \pi^2) (4 \pi)^4} \,{\rm Tr}\, (T^A T^B) \,{\rm
Tr}\, [ P_+ \gamma^k \gamma^m \gamma^l \gamma^n \gamma^p \gamma^q] \times \\
\int d^3y_3 \, d^3y_4 \, {1 \over |y_{34}|}\,  \partial_m {1\over |y_{12}|} \, 
\partial_n {1 \over |y_{23}|} \, \partial_p { 1 \over |y_{34}|^2} \, 
\partial_q {1 \over |y_{41}|} \,.
\nonumber
\end{eqnarray}
Using differential regularization we first confront the subdivergence
which occurs in the $\lambda / q$ loop.  We start by writing
\begin{eqnarray}
\label{Regulate}
{1 \over |y_{34}|}\, \partial_p { 1 \over |y_{34}|^2} = {2 \over 3}
\partial_p {1 \over |y_{34}|^3} \,.
\end{eqnarray}
This would lead to a divergence in the integrals over $y_3$ and $y_4$. 
In differential regularization this is handled by means of the substitution
\begin{eqnarray}
\label{drid}
{1 \over |y_{34}|^3} = - \partial_p \nabla^2 
{\log M |y_{34}| \over |y_{34}|} \,,
\end{eqnarray}
where $\nabla^2$ is the 3D Laplacian, and $M$ is an arbitrary mass
unit.  It is straightforward to show that, after insertion of
(\ref{Regulate}), (\ref{drid}), the integrals in (\ref{jjcorr}) can be
done using the delta functions which appear after partial integration.
For our purposes, it is sufficient to focus on the divergent
($M$-dependent) part.  To isolate this we apply $M \partial/\partial
M$. The regulated subdivergence (\ref{Regulate}) (which contains all
the $M$-dependence) then becomes
\begin{eqnarray}
- {2 \over 3} \partial_p \nabla^2 {1 \over |y_{34}|} = 
{2 \over 3}\, (4 \pi)\, \partial_p \delta(y_{34}) \,. 
\end{eqnarray}
Substituting back into $\langle j j \rangle$, we find
\begin{eqnarray}
M {\partial \over \partial M} \langle j^{Ak}(y_1) \, j^{Bl}(y_2) \rangle  =
 -{ 16 \,g^2 \over 3 (4 \pi^2) (4 \pi)^3} \,{\rm Tr}\, (T^A T^B)
\,{\rm Tr}\, 
[ P_+ \gamma^k \gamma^m \gamma^l \gamma^n \gamma^p \gamma^q] \times \\
\partial_m {1\over |y_{12}|}
\partial_n \partial_p \partial_q \int d^3y_3 \, {1 \over |y_{23}|} 
{1 \over |y_{31}|} \,,
\nonumber
\end{eqnarray}
where all the derivatives act on the $2$ coordinate.

Using the $\gamma$-matrix identity $\gamma^n \gamma^p \gamma^q =
\gamma^n \gamma^{pq} + \gamma^n g^{pq}$, we generate a 3D Laplacian;
acting on $1/|y_{23}|$ this generates another $\delta$-function which
allows us trivially to do the integral.  We finally obtain
\begin{eqnarray}
\label{mdm}
\nonumber
M {\partial \over \partial M} \langle j^{Ak}(y_1) \, j^{Bl}(y_2) \rangle  &=&
 -{ 16 \,g^2 \over 3 (4 \pi^2) (4 \pi)^2} \,{\rm Tr}\, (T^A T^B) 
\,{\rm Tr}\, [ P_+ \gamma^k \gamma^m \gamma^l \gamma^n]
\partial_m  {1 \over |y_{12}|} \partial_n {1 \over |y_{12}|}\,, \\
&=&  { 16 \,g^2 \over 3 (4 \pi^2) (4
\pi)^2} 
\,{\rm Tr}\, (T^A T^B)  \left( {2 \over |y_{12}|^4} J^{kl}(y_{12}) 
\right) \,,
\label{33Final}
\end{eqnarray}
where $J^{kl}$ is the
conformal Jacobian tensor $J^{kl}(y) \equiv \delta^{kl} - 2 y^k y^l / y^2$.

We may use the result (\ref{33Final}) to the calculate the anomalous
dimension of $j^k$.  Because $j^k$ is a conformal primary defect
vector operator of some scaling dimension $\Delta$, 
its two-point function must have the form
\begin{eqnarray}
\label{confcorr}
 \langle j^{Ak}(y_1) \, j^{Bl}(y_2) \rangle  = - c(g) \, {\rm Tr}\, 
(T^A T^B)\, {M^4  \over
 8 \pi^2 |M y_{12}|^{2\Delta} } \, J^{kl}(y_{12})\,.
\end{eqnarray}
We have normalized the correlator to the free-field value at $g=0$.
To order $g^2$, we expect that the scaling dimension is $\Delta = 2 +
\gamma(g)$ and $c(g) = 1 +g^2 a$ for some constant $a$, 
where $\gamma(g)$ is the anomalous dimension of $j$.
We may determine $\gamma$ using the perturbative relation
\begin{eqnarray}
{M^4 \over |My_{12}|^{2\Delta}} \sim {1 \over |y_{12}|^4} [1 -2 \gamma(g) 
\log |y_{12}M|] \,.
\end{eqnarray}
We insert this in (\ref{confcorr}), compute the scale derivative $M
\partial / \partial M$, and compare with (\ref{33Final}). We thus
identify the anomalous dimension to order $g^2$ as 
\begin{eqnarray}
\gamma(g^2) = { 2 g^2 \over 3 \pi^2} \,.
\end{eqnarray}
Note that it is positive, as required by unitarity.

A term of the form (\ref{confcorr}) appearing in $\langle J J \rangle$
seems to be inconsistent with the conservation of $J$ when $\Delta
\neq 2$, so the $M$-dependence coming from this term has to cancel out
in the full $\langle J J \rangle$ correlator. Let us now show that it
is indeed canceled by the contributions from $\langle {\cal J} j
\rangle$ and $\langle j {\cal J} \rangle$.  We have
\begin{eqnarray}
\langle {\cal J}^{Ak}(z_1) \, j^{Bl}(y_2) \rangle  =
 -{ 4 \,g^2 \over (4 \pi^2)^2 (4 \pi)^3} \,{\rm Tr}\, (T^A T^B) 
{\rm Tr}\, [ \gamma^\mu \, \gamma^k \, 
\gamma^\nu \, \gamma^m P_+ \gamma^l P_+ \,\gamma^n] \times \\
\int d^3y_3 \, d^3y_4 \, {1 \over |y_{34}|} (\partial_\mu {1 \over z_{31}^2}) 
(\partial_\nu {1 \over z_{14}^2}) (\partial_m {1 \over |y_{42}|}) 
(\partial_n {1 \over |y_{23}|}) \,. 
\nonumber
\end{eqnarray}
Again we must regulate the subdivergence, which in this case comes
from the divergent $\lambda/\lambda/q$ loop.  Note that there is no
overlapping divergence, since the limit where the $q$ propagator
approaches the $\bar\Psi \Psi$ vertex is finite.  The subgraph is
\begin{eqnarray}
(P_+ \gamma^\mu \gamma^k \gamma^\nu P_+)\, {1 \over |y_{34}|}
(\partial_\mu {1 \over z_{31}^2}) (\partial_\nu {1 \over z_{14}^2}) \,.
\end{eqnarray}
We are using a convention where the derivative always acts on the
first coordinate listed in a propagator.  Switching the $\mu$-derivative 
to the 1 coordinate, we can pull it out to find
\begin{eqnarray}
(P_+ \gamma^\mu \gamma^k \gamma^\nu P_+) \left[ \partial^1_\mu \left(
{1 \over |y_{34}|} {1 \over z_{31}^2} (\partial_\nu {1 \over
z_{14}^2}) \right) + {1 \over |y_{34}|} {1 \over z_{31}^2}
(\partial_\mu \partial_\nu {1 \over z_{14}^2}) \right] \equiv 
 (P_+ \gamma^\mu \gamma^k \gamma^\nu P_+) \left[ \partial^1_\mu 
B_\nu + C_{\mu\nu} \right].
\end{eqnarray}
For the first term, the vector $B_\nu$ is log divergent by power
counting, but since it is a vector, a further derivative can be
extracted and what remains is finite.  Hence the divergence lies in
the second piece $C_{\mu\nu}$, on which we now concentrate.

In differential regularization, tensors like $\partial_\mu
\partial_\nu$ are split into their trace and traceless parts; the
divergence lies entirely in the trace.  In our case, the situation is
slightly more subtle due to the defect.  Since $y_3$ and $y_4$ are
only three-dimensional points, the traceless tensor that is
appropriate is the three-dimensional one $\partial_k \partial_l -
\tf13 \delta_{kl} \nabla^2$. Thus, we are forced to split the tensor
$C_{\mu\nu}$ into defect indices and transverse indices separately.

If one of $\mu$ and $\nu$ is a defect coordinate while the other is not,
the projection matrices $P_+$ annihilate one another.  Thus we get
\begin{eqnarray}
(P_+ \gamma^\mu \gamma^k \gamma^\nu P_+) \, C_{\mu\nu} = {1 \over
|y_{34}|} ({1 \over z_{13}^2}) \left[ (P_+ \gamma^m \gamma^k \gamma^n
) \partial^1_m \partial^1_n + (P_+ \gamma^x \gamma^k \gamma^x
) (\partial_x)^2 \right] ({1 \over z_{14}^2}) \,.
\end{eqnarray}
We split $\partial_m \partial_n = \tf13 \delta_{mn} \nabla^2 +$
traceless, and ignore the traceless part.  Using $\gamma^x \gamma^k
\gamma^x = - \gamma^k$, $\gamma^m \gamma^k \gamma^n \delta_{mn} = -
\gamma^k$, and $(\partial_3)^2 = \square - \nabla^2$ with $\square$
the 4D Laplacian, we arrive at
\begin{eqnarray}
\label{43Intermediate}
(\gamma^k P_+) \, {1 \over |y_{34}|} ({1 \over z_{13}^2}) 
( \tf23 \nabla^2 - \square) ({1 \over z_{14}^2}) \,.
\end{eqnarray}
To treat the $\nabla^2$ term, we first switch the 3D Laplacian to act
on the 4 coordinate and then integrate by parts.  Terms with an
overall $\nabla_k$ are not singular, so the relevant term involves
$\nabla^2 (1 / |y_{43}|) = - 4 \pi \delta(y_{43})$.  With this
$\delta$-function present, we can combine the two 4D propagators and
regulate the resulting expression in the standard 4D fashion
\begin{eqnarray}
({1 \over z_{13}^2}) ({1 \over z_{14}^2}) \rightarrow ({1 \over
z_{13}^4}) \rightarrow - \tf14 \square { \log M^2 z_{13}^2 \over
z_{13}^2} \,.
\end{eqnarray}
Now turn to the $\square$ term in (\ref{43Intermediate}).  The 4D
Laplacian acts on the propagator to produce $\delta(z_{14})$, which
allows us to regulate
\begin{eqnarray}
 {1 \over |y_{34}|} {1 \over z_{13}^2} \rightarrow {1 \over |y_{34}|^3} 
\rightarrow - \nabla^2 { \log M |y_{34}| \over |y_{34}|} \,.
\end{eqnarray}
The total expression for both terms is hence
\begin{eqnarray}
 (\gamma^k P_+) \, \left[ \tf16 (4 \pi)  \left(\square { \log M^2 z_{13}^2 
\over z_{13}^2} \right) \delta(y_{43}) -  (4 \pi^2) \nabla^2 
{ \log M |y_{34}| \over |y_{34}|} \delta(z_{14})\right] \,.
\end{eqnarray}
Taking $M \partial/ \partial M$ to isolate the $M$-dependent part of
the subdivergence, we have
\begin{eqnarray}
\tf23  (\gamma^k P_+) \, (4 \pi^2) (4 \pi) \delta(z_{14}) \delta(y_{34}) \,.
\end{eqnarray}

Substituting the subdivergence back into the total expression for
$\langle {\cal J} j \rangle$, and recalling that all the terms we dropped 
are $M$-independent, we thus have 
\begin{eqnarray}
\label{43Final}
M {\partial \over \partial M} \langle {\cal J}^{Ak}(z_1) \, 
j^{Bl}(y_2) \rangle  =
 -{ 8 \,g^2 \over 3 (4 \pi^2) (4 \pi)^2} \,{\rm Tr}\, (T^A T^B) 
{\rm Tr}\, [ P_+ \gamma^k \, \gamma^m \, \gamma^l \,  \,\gamma^n] \times \\
\int d^3y_3 \, d^3y_4 \, \delta(z_{13}) \delta(y_{34}) (\partial_m 
{1 \over |y_{42}|}) (\partial_n {1 \over |y_{23}|}) \nonumber \\
= { 8 \,g^2 \, \delta(x_1) \over 3 (4 \pi^2) (4 \pi)^2} 
\,{\rm Tr}\, (T^A T^B) 
{\rm Tr}\, [ P_+ \gamma^k \, \gamma^m \, \gamma^l \,  \,\gamma^n] 
(\partial_m {1 \over |y_{12}|}) (\partial_n {1 \over |y_{12}|})\,.
\nonumber
\end{eqnarray}
The contribution from $\langle j {\cal J} \rangle$ is identical up to
exchanging $x_1 \leftrightarrow x_2$.  Consequently from
(\ref{33Final}) and (\ref{43Final}) we see
\begin{eqnarray}
M {\partial \over \partial M} \left( \delta(x_2) \langle {\cal J}^{Ak}(z_1) \, 
j^{Bl}(y_2) \rangle + \delta(x_1) \langle j^{Ak}(y_1) \, {\cal J}^{Bl}(z_2) 
\rangle + \delta(x_1) \delta(x_2) \langle j^{Ak}(y_1) \, j^{Bl}(y_2) 
\rangle \right) = 0,
\end{eqnarray}
and $\vev{J_V J_V}$ has no divergences, as required.  Notice that we
had to isolate the $M$-dependent part of $\langle {\cal J} j \rangle$
in order to obtain the $\delta(x_1)$.  The complete correlation
function $\langle {\cal J}(z_1) j(y_2) \rangle$ is not localized on
the defect in the $x_1$ coordinate, only the divergent part is.

\section{Anomalous Dimensions from Gravity}

\label{ads_section}

As discussed in section \ref{bdod}, the reduced conformal group
$SO(3,2)$ preserved by the defect is more permissive than the full
group $SO(4,2)$, but still places strong constraints on correlation
functions (in this section we specialize to the case of $d=4$, though
the generalization to arbitrary $d$ should be straightforward).  The
simplest permitted correlators involving 4D operators are the
one-point function $\langle {\cal O}_4 \rangle$ and the mixed
two-point function $\langle {\cal O}_4 {\cal O}_3 \rangle$.  As with
two- and three-point functions in ordinary CFT, the coordinate
dependence of these correlators is completely specified by the
symmetry.  It was shown in \cite{dfo} that the corresponding $AdS_4$
interactions $\int_{AdS_4} \phi_5$ and $\int_{AdS_4} \phi_5 \psi_4$
lead to dCFT correlation functions with the correct structure.

In this section we consider the gravity calculation of two-point
correlation functions between four-dimensional operators.  Just as the
one-point and mixed two-point cases in dCFT are somewhat analogous to the two-
and three-point functions of ordinary CFT, the case of ambient
two-point functions is analogous to the ordinary CFT four-point
function. As discussed in section \ref{bdod} it is the simplest
correlator admitting a nontrivial coordinate dependence, and this
coordinate dependence encodes the dimensions of operators living on
the defect.

\begin{figure}
 \centerline{\psfig{figure=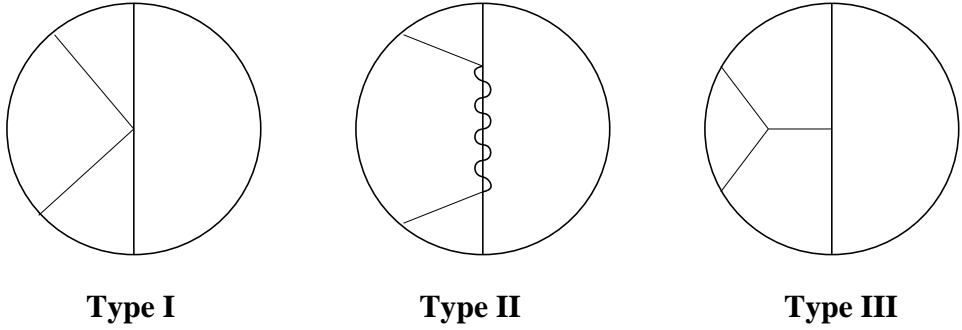,width=5.0in}}
 \caption{Diagrams appearing in the computation of 
$\vev{{\cal O}_4 {\cal O}_4}$. The circle
 represents the boundary of $AdS_5$, where we insert the operators,
 and the vertical straight line crossing it represents the $AdS_4$ brane.
The other solid lines are bulk scalar propagators, and the curly line is
a brane scalar propagator.}
\label{witten}
  \end{figure}

There are a number of ways to generate contributions to these
correlators in the gravity theory with a probe brane. At the leading
order in the interaction with the brane (which in the D3/D5 system is
the same as leading order in $g_s M$, a disc diagram in string theory),
we encounter three classes of diagrams in perturbative calculations,
which are drawn in figure \ref{witten}.  Each diagram is independent
of the others so it must individually lead to a contribution with the
form (\ref{ofourofour}) to ${\cal M}_{\Delta_1 \Delta_2}(\vec{y}_1,
x_1;\vec{y}_2, x_2) \equiv \langle {\cal O}_{\Delta_1}(\vec{y}_1,x_1)
\, {\cal O}_{\Delta_2}(\vec{y}_2, x_2) \rangle$.

Below we present the general analysis for the various types of
diagrams.  We will only consider the two-point functions of scalars,
interacting with other bulk scalars as well as brane scalars; generically
intermediate higher spin fields would also contribute to the two-point
function of scalar operators but we do not consider them here.  We
adapt the method of ``without really trying'' \cite{wrt} that
was created for evaluating CFT four-point functions.

We work in Euclidean $AdS$, with quadratic terms normalized as 
\begin{eqnarray}
{\cal L} &=&{1 \over 2 \kappa_5^2} \int d^5z \sqrt{-g_5} \left( -
{\cal R}_5 + {1 \over 2} \sum_i (\partial_\mu \phi_i \partial^\mu
\phi_i + m_i^2 \phi_i \phi_i)\right) + \\ && {1 \over 2 \kappa_4^2}
\int d^4z \sqrt{-g_4} \left( {1 \over 2} \sum_j (\partial_{\mu'}
\psi_j \partial^{\mu'} \psi_j + m_j^2 \psi_j \psi_j) \right) \,.
\nonumber
\end{eqnarray}
The Euclidean $AdS_5$ metric is
\begin{eqnarray}
ds^2 = {1 \over z_0^2} \left( dz_0^2 + d\vec{y}^2 + dx^2 \right) \,,
\end{eqnarray}
where we have set the $AdS$ curvature radius to one (it is easy to
reinstate it if desired), and the $AdS_4$ metric is the same without
the $dx^2$ term. From here on we will use the letters $z$ and $w$ to
denote the $AdS$ coordinates, and $d^4z$, $d^5z$ will be the volume
elements of $AdS_4$ and $AdS_5$, respectively.

\subsection{Type I}
\label{TypeISec}

The type I diagrams are the simplest, representing contact terms that
involve only one integral over an internal point.  Moreover, the more
complicated diagrams of type II and type III can be shown to reduce to
type I diagrams.  We shall therefore begin with them.

Assuming that the brane action includes a coupling of bulk fields of the form
\begin{eqnarray}
\lambda_I \int d^4z \sqrt{-g_4} \, \phi_1 \phi_2 \,, 
\end{eqnarray}
we find\footnote{Assuming that $\phi_1$ and $\phi_2$ are different
fields; an additional factor of 2 arises if they are the same.} a
contribution to the two-point function ${\cal M}_{\Delta_1 \Delta_2}$
of the form
\begin{eqnarray}
{\cal M}_{\Delta_1 \Delta_2}= - \lambda_I \int d^4z \sqrt{-g_4}
\, K_{\Delta_1}(z;\vec{y}_1,x_1) K_{\Delta_2}(z;\vec{y}_2,x_2) \,,
\end{eqnarray}
where $K_\Delta$ is the standard bulk-to-boundary propagator:
\begin{eqnarray}
K_{\Delta}(z;\vec{y},x) = C_\Delta \left ( \frac{z_0}{x^2+(\vec{y} - 
\vec{z})^2 + z_0^2} \right )^\Delta \equiv C_\Delta 
\tilde{K}_\Delta(z;\vec{y},x)\,,
\end{eqnarray}
with $C_\Delta = \Gamma(\Delta) / (\pi^2 \Gamma(\Delta - 2))$.
The contribution to the two-point function is thus:
\begin{eqnarray}
{\cal M}_{\Delta_1\Delta_2}(\vec{y}_1, x_1;\vec{y}_2, x_2) = -
\lambda_I C_{\Delta_1} C_{\Delta_2} I_{\Delta_1
\Delta_2}(\vec{y}_1,x_1;\vec{y}_2,x_2) \,,
\end{eqnarray}
where
\begin{eqnarray}
\label{zintegral}
I_{\Delta_1\Delta_2}(\vec{y}_1,x_1;\vec{y}_2,x_2) =
\int \frac{dz_0 d \vec{z}}{z_0^4}
\left ( \frac{z_0}{x_1^2+(\vec{y}_1 - \vec{z})^2 + z_0^2} \right )^{\Delta_1}
\left ( \frac{z_0}{x_2^2+(\vec{y}_2 - \vec{z})^2 + z_0^2} \right )^{\Delta_2}.
\end{eqnarray}
This will be the standard object, in terms of which we will express the
more complicated diagrams.  We use translation invariance along the
defect to set $\vec{y}_2=0$, and denote $\vec{y} \equiv \vec{y}_1$.
The integral can be performed using Feynman parameters, giving
\begin{eqnarray}
\label{Iintegral}
I_{\Delta_1 \Delta_2}(\vec{y},x_1;0,x_2) = D_{\Delta_1 \Delta_2} \int_0^1 da
{a^{\Delta_1-1} (1-a)^{\Delta_2-1} \over (ax_1^2 + (1-a) x_2^2 + a(1-a)
\vec{y}^2)^{(\Delta_1+\Delta_2)/2}} \,,
\end{eqnarray}
where 
\begin{eqnarray} 
D_{\Delta_1\Delta_2} \equiv {\pi^{3/2} \over 2} {\Gamma\left( {\Delta_1 + 
\Delta_2 -3 \over 2} \right) \Gamma\left( {\Delta_1 + \Delta_2 \over 2} 
\right) \over\Gamma(\Delta_1) \Gamma(\Delta_2)} \,.
\end{eqnarray}

Based on the discussion in section \ref{bdod} we expect the result of
this integral to have the form $I_{\Delta_1 \Delta_2} = f(\xi) /
x_1^{\Delta_1} x_2^{\Delta_2}$. This implies that if we define
$f(\vec{y},x_1,x_2) = x_1^{\Delta_1} x_2^{\Delta_2} I_{\Delta_1
\Delta_2}(\vec{y},x_1;0,x_2)$, it should be purely a function of $\xi$, 
so it should obey
\begin{eqnarray}
(x_1 {\del \over \del x_1} + x_2 {\del \over \del x_2} + 
2 \vec{y}^2 {\del \over \del \vec{y}^2}) f(\vec{y},x_1,x_2) = 0
\end{eqnarray}
and
\begin{eqnarray}
(x_1 {\del \over \del x_1} - x_2 {\del \over \del x_2} - 
2 (x_1^2 - x_2^2) {\del \over \del \vec{y}^2}) f(\vec{y},x_1,x_2) = 0.
\end{eqnarray}
By simple manipulations on the integral (\ref{Iintegral}) it is easy
to show that these equations are indeed satisfied, so the expressions
that we find here are consistent with the form (\ref{ofourofour}) of
the two-point function, dictated by conformal invariance.

We may then solve (\ref{Iintegral}) by simply setting
$\vec{y}=0$ and restoring the full $\xi$-dependence at the end.  We find
\begin{eqnarray}
\nonumber
I_{\Delta_1 \Delta_2}(\xi,x_1,x_2) &=& {\pi^{3/2} \Gamma\left({\Delta_1 + 
\Delta_2 - 3 \over 2}\right) \Gamma\left({\Delta_1 + \Delta_2 \over 2}\right) 
\over 2 x_1^{\Delta_1} x_2^{\Delta_2} \Gamma\left(\Delta_1 + \Delta_2 
\right)} \, \zeta^{\Delta_1}\, \cdot \,
_2F_1\left( {\Delta_1 + \Delta_2 \over 2}, \Delta_1; \Delta_1 + \Delta_2; 
1 - \zeta^2 \right) \,, \\
\zeta &\equiv& 1 + 2 \xi + 2 \sqrt{\xi(\xi+1)} \,,
\label{Isoln}
\end{eqnarray}
valid for $\xi > 0$.
Although this is not manifestly symmetric between $\Delta_1$ and
$\Delta_2$, it can be shown to be so using hypergeometric identities.
The small-$\xi$ expansion of (\ref{Isoln}) is simply
\begin{eqnarray}
I_{\Delta_1 \Delta_2}(\xi\rightarrow 0, x_1, x_2) = {\pi^{3/2}
\Gamma\left({\Delta_1 + \Delta_2 - 3 \over 2}\right)
\Gamma\left({\Delta_1 + \Delta_2 \over 2}\right) \over 2
x_1^{\Delta_1} x_2^{\Delta_2} \Gamma\left(\Delta_1 + \Delta_2 \right)}
\, \left( 1 + \co(\xi) \right) \,.
\end{eqnarray}
For odd $\Delta_1 + \Delta_2$, (\ref{Isoln}) reduces to nice rational
functions of $\xi$, while for even $\Delta_1 + \Delta_2$ there is a
logarithmic term, and the leading large-$\xi$ behavior of $I_{\Delta_1
\Delta_2}$ is proportional to $\log (\xi) / x_1^{\Delta_1}
x_2^{\Delta_2} \xi^{\max(\Delta_1,\Delta_2)}$.  In subsection
\ref{CorrAnomDimSec} we explain these logarithms as resulting from the
anomalous dimensions of the reduced operators appearing in the BOPE of
$\co_{\Delta_1}$ and $\co_{\Delta_2}$.

As a simple example, $I_{1 \, 1}(\xi, x_1, x_2)$ may be evaluated using 
either (\ref{Iintegral}) or (\ref{Isoln}), giving
\begin{eqnarray}
\label{ioneone}
I_{1 \, 1}(\xi,x_1,x_2)  = -{\pi^2 \over {4 x_1 x_2 \sqrt{\xi (\xi+1)}}} 
\log \left( {{\xi+{1\over
2}+\sqrt{\xi(\xi+1)}} \over {\xi+{1\over 2}-\sqrt{\xi(\xi+1)}}} \right) \,.
\end{eqnarray}
Note that for $\Delta_1=\Delta_2=1$ the integral (\ref{zintegral})
actually diverges, but we can still assign to it the above value
(following from (\ref{Iintegral})) by analytic continuation in
$\Delta_1$ and $\Delta_2$. For $\Delta_1+\Delta_2=3$ there is a pole in this
analytically continued expression, but it is well-behaved for
$\Delta_1+\Delta_2 > 3$. As another example, we have
\begin{eqnarray}
\label{itwothree}
I_{3\, 2}(\vec{y},x_1;0,x_2) = {\pi^2 \over {(2x_1)^3 (2x_2)^2
2 (\xi+1)^2}}\,.
\end{eqnarray}
Note that $-1 < \xi < 0$ does not occur.

Instead of computing them directly, we can 
obtain all other $I_{\Delta_1 \Delta_2}$ with integer
$\Delta_1$ and $\Delta_2$ by taking derivatives of (\ref{ioneone}) and
(\ref{itwothree}).  From the expression (\ref{Iintegral}) we obtain
the relation
\begin{eqnarray}
\label{irecursive}
{\partial \over {\partial \vec{y}^2}} I_{\Delta_1\Delta_2} =
{1 \over 4 x_1 x_2} {\partial \over \partial \xi} I_{\Delta_1\Delta_2} =
-\left({2\Delta_1\Delta_2 \over \Delta_1+\Delta_2-3}\right)
I_{\Delta_1+1\, \Delta_2+1} \,.
\end{eqnarray}
Furthermore, defining $\tilde{x}^2 \equiv \vec{y}^2 + x_1^2$ and
holding it fixed, we can also obtain
\begin{eqnarray}
\label{irecursive2}
{\partial \over {\partial \vec{y}^2}} \Bigg|_{\tilde{x}^2}
I_{\Delta_1\Delta_2} = {2\Delta_1(\Delta_1 + 1) \over \Delta_1+\Delta_2-3}
I_{\Delta_1+2, \Delta_2} \,.
\end{eqnarray}
As we shall see, all the other contributions to two-point functions at leading 
order can be written in terms of the functions $I_{\Delta_1 \Delta_2}$.

It is worth noting that couplings with derivatives can also appear in
the action on the brane.  For example, consider a term of the form
\begin{eqnarray}
\varsigma_I \int d^4z \sqrt{-g_4} \, \phi_1 (z_0 \partial_{z_3} \phi_2) \,, 
\end{eqnarray}
where the brane is located at $z_3 = 0$ and
the restriction of the derivative of $\phi_2$ normal to the brane appears in
the interaction.  This leads to the contribution
\begin{eqnarray}
{\cal M}_{\Delta_1\Delta_2}(\vec{y}_1, x_1;\vec{y}_2, x_2)
= - \varsigma_I \int d^4z \sqrt{-g_4}
\, K_{\Delta_1}(z;\vec{y}_1,x_1) \left(z_0 \partial_{z_3} 
K_{\Delta_2}(z;\vec{y}_2,x_2)\right) \,,
\end{eqnarray}
where we set $z_3 = 0$ for $K_{\Delta_2}$ only after acting with the 
derivative.  We find
\begin{eqnarray}
z_0 \partial_{z_3} \tilde{K}_\Delta(z;\vec{y},x) = 2 x \Delta 
\tilde{K}_{\Delta+1}(z;\vec{y},x) \,,
\end{eqnarray}
and correspondingly
\begin{eqnarray}
\label{TypeIDerivResult}
{\cal M}_{\Delta_1\Delta_2}(\vec{y}_1, x_1;\vec{y}_2, x_2) = - 2 
\varsigma_I  \Delta_2
C_{\Delta_1} C_{\Delta_2} [ x_2 I_{\Delta_1
\Delta_2+1}(\vec{y}_1,x_1;\vec{y_2},x_2) ]\,.
\end{eqnarray}
From the discussion above it is clear that $x_2 I_{\Delta_1
\Delta_2+1}(\vec{y}_1,x_1;\vec{y_2},x_2)$ has the correct coordinate
dependence to satisfy the conformal invariance condition (\ref{ofourofour}).

\subsection{Type II}

We assume a brane interaction between bulk fields $\phi_i$, $i=1,2$
with associated conformal dimensions $\Delta_i$, and a brane field $\psi$ with
associated conformal dimension $\Delta$, of the form
\begin{eqnarray}
\sum_{i=1}^2 \lambda^i_{II} \int d^4z \sqrt{-g_4} \, \phi_i \psi \,. 
\end{eqnarray}
The contribution to a two-point function is then
\begin{eqnarray}
{\cal M}_{\Delta_1 \Delta_2} &=& \lambda^1_{II} \lambda^2_{II} (2
\kappa_4^2) C_{\Delta_1} C_{\Delta_2} \int {d^4w \over w_0^4}
K_{\Delta_2}(w; \vec{y}_2, x_2) B(w; \vec{y}_1, x_1) \,, \\
B(w; \vec{y}_1, x_1) &\equiv& \int {d^4z \over z_0^4} \, 
K_{\Delta_1}(z; \vec{y}_1, x_1) \, G_\Delta(u) \,, 
\end{eqnarray}
where $G_\Delta(u)$ is a scalar bulk-to-bulk propagator for the field $\psi$,
depending on the chordal distance between the points $z$ and $w$,
\begin{eqnarray}
u = {1 \over 2 z_0 w_0} \left( (\vec{z} - \vec{w})^2  + 
(z_0 - w_0)^2 \right)  \,.
\end{eqnarray}
Note that the points $z$, $w$ are both on the $AdS_4$ brane.

We can use a method analogous to the ``not really trying'' method of 
\cite{wrt} to determine
$B(w; \vec{y}_1, x_1)$; the idea is to use symmetries and the scalar
wave operator to obtain an ordinary differential equation for $B$.
Poincar\'e invariance requires that $B$ depend only on $z_0$, $x_1$
and $|\vec{z} - \vec{y}_1|$.  Scaling all coordinates by a constant
$\Lambda$, $u$ is invariant and hence
$B \rightarrow \Lambda^{-\Delta_1} B$, implying
\begin{eqnarray}
B(w; \vec{y}_1, x_1)  = x_1^{-\Delta_1} J\left(\eta, {z_0 \over x_1} \right)\,,
\end{eqnarray}
where
\begin{eqnarray}
\eta \equiv {z_0 x_1 \over z_0^2 + x_1^2 + |\vec{z} - \vec{y}_1|^2} \,
\end{eqnarray}
is a scale-invariant combination of the variables.  One can also check
that $\eta$ is invariant under an inversion of all coordinates,
\begin{eqnarray}
z_\mu \rightarrow {z_\mu \over z^2} \,, \quad \quad 
x_1 \rightarrow {x_1 \over x_1^2 + y_1^2} \,, \quad \quad 
\vec{y}_1 \rightarrow {\vec{y}_1 \over x_1^2 + y_1^2} \,.
\end{eqnarray}
Under this inversion one finds that $B$ and $x_1^{-\Delta_1}$
transform identically, meaning $J$ is invariant; since $z_0/x_1$ is
not inversion-invariant, we must have simply
\begin{eqnarray}
\label{BExpression}
B(w; \vec{y}_1, x_1)  = x_1^{-\Delta_1} J\left(\eta \right)\,.
\end{eqnarray}
Applying the $AdS_4$ wave operator obeying $(- \square + m_\psi^2) \,
G_\Delta(u) = \delta(z,w)$ to (\ref{BExpression}), we obtain the ODE for $J$,
\begin{eqnarray}
\left[ (4 \eta^4 - \eta^2) \partial_\eta^2 + (8 \eta^3 + (d-1) \eta) 
\partial_\eta + m^2_\psi \right] J(\eta) = \eta^{\Delta_1} \,.
\end{eqnarray}
Looking for a power series solution of the form
\begin{eqnarray}
J(\eta) = \sum_k  b_k \, \eta^k\,,
\end{eqnarray}
we find
\begin{eqnarray}
b_{\Delta_1} &=& 0 \,, \\
b_{\Delta_1 -2} &=& {1 \over 4 (\Delta_1 -1)(\Delta_1 - 2)} \,, \\
b_{k-2} &=& { (k-\Delta) (k + \Delta - d) \over 4 (k-1)(k-2)} \, b_k \,, \quad 
k = \Delta_1-2, \Delta_1-4, \ldots, k_{min},
\end{eqnarray}
with $k_{min} = \Delta$ and other $b_k = 0$.  
The total 
contribution is then
\begin{eqnarray}
{\cal M}_{\Delta_1 \Delta_2} &=& \lambda^1_{II} \lambda^2_{II} (2
\kappa_4^2) C_{\Delta_1} C_{\Delta_2} \left\{ \sum_k b_k \, x_1^{k-\Delta_1}
I_{k,\Delta_2}(\vec{y}_1,x_1;\vec{y}_2,x_2) \right\}\,.
\end{eqnarray}
The amplitude has been reduced to a sum of diagrams of type I
multiplied by a function of coordinates.  Each term has the
correct form (\ref{ofourofour}) dictated by defect conformal
invariance.  This solution requires $\Delta_1 - \Delta$ to be a
positive even integer; a similar constraint appeared in \cite{wrt}. We
do not know how to compute other type II diagrams where this
constraint is not satisfied.  Of course, we can repeat the same
analysis exchanging $\phi_1$ and $\phi_2$, and if both
$\Delta_1-\Delta$ and $\Delta_2-\Delta$ are even integers we obtain
the same answer in different forms.

\subsection{Type III}

Assuming a bulk interaction of the form
\begin{eqnarray}
\label{BulkInt}
S_{bulk} = {1 \over 2 \kappa_5^2} \int d^5z \sqrt{-g_5} \left( 
\phi_{\Delta_1} \phi_{\Delta_2} \phi_{\Delta_3} \right) \,,
\end{eqnarray}
and a brane interaction
\begin{eqnarray}
\label{BraneInt}
S_{brane} = \lambda \int d^4z \sqrt{-g_4} \, \phi_{\Delta_3} \,,
\end{eqnarray}
the usual manipulations result in the expression
\begin{eqnarray}
{\cal M}_{\Delta_1 \Delta_2}(\vec{x}_1, y_1; \vec{x}_2, y_2) &=& \lambda \int 
{d^4w \over w_0^4} \, A(w;
\vec{y}_1, x_1; \vec{y}_2, x_2) \,, \\ A(w; \vec{y}_1,
x_1; \vec{y}_2, x_2) &\equiv& \int {d^5z \over
z_0^5} \, K_{\Delta_1}(z; \vec{y}_1, x_1)
K_{\Delta_2}(z; \vec{y}_2, x_2) \, G_{\Delta_3}(u)
\,, \nonumber
\end{eqnarray}
where $G_\Delta$ is now an $AdS_5$ bulk-to-bulk propagator, and the
chordal distance $u$ now reflects the fact that $z$ is not pinned to the brane,
\begin{eqnarray}
u = {1 \over 2 z_0 w_0} \left( (\vec{z} - \vec{w})^2 + (z_3)^2 + 
(z_0 - w_0)^2 \right)  \,.
\end{eqnarray}
Note that the propagator generates a factor $2 \kappa_5^2$ that
cancels the factor from the coupling.  We notice immediately that the
function $A(w;1;2)$ is identical to the function $A$ defined in (2.11)
of \cite{wrt}.  It involves integrating a point over the complete
bulk, and is not aware of the presence of the defect, other than the
fact that $w_3 = 0$.  Hence one can take the results from that
analysis wholesale.  We proceed by shifting $\vec{y}_1 \rightarrow
0$ and inverting the coordinates into primed coordinates, and we arrive at
\cite{wrt}
\begin{eqnarray}
A(w;1;2) &=& {1 \over (\vec{x}_{12} + y_{12})^{2 \Delta_2}} \,
I(\vec{w}'- \vec{y}_{12}', w_3' - x_{12}', w_0') \,,\\
I(\vec{y},x,w_0) &=& w_0^{\Delta_{12}} \sum_k a_k \left( w_0^2 \over w_0^2 +
\vec{y}^2 + x^2 \right)^k\,,
\end{eqnarray}
where the $a_k$ are given recursively by
\begin{eqnarray}
\label{aCoeff}
a_k &=& 0 \quad {\rm for} \: k \geq \Delta_2 \,, \\ a_{\Delta_2 - 1}
&=& {1 \over 4 (\Delta_1 - 1)(\Delta_2 - 1)} \,,\\ a_{k-1} &=& { (k -
\tf{\Delta_3}{2} + \tf{\Delta_{12}}{2})(k - 2 + \tf{\Delta_3}{2} +
\tf{\Delta_{12}}{2}) \over (k-1)(k-1 + \Delta_{12})} \, a_k \,.
\end{eqnarray}
This is the basic result of ``not really trying''.  The series terminates
below at $k = (\Delta_3 - \Delta_{12})/2$ if $\Delta_1 + \Delta_2 -
\Delta_3$ is a positive even integer. Again, we do not know how to
evaluate the diagram if this constraint is not satisfied. However, it
seems that the constraint is always satisfied in type IIB supergravity
on $AdS_5\times S^5$ (though it is not satisfied in the full string
theory, nor in other backgrounds).

Assembling the total result for the diagram, including undoing the 
inversion and translation, one arrives at
\begin{eqnarray}
{\cal M}_{\Delta_1 \Delta_2}(\vec{y}_1, x_1; \vec{y}_2, x_2) = 
\lambda C_{\Delta_1}
C_{\Delta_2} \sum_k a_k (\vec{y}_{12}^2 +
x_{12}^2)^{k-\Delta_2 } I_{\Delta_1-\Delta_2+k,k}(\vec{y}_1,x_1;
\vec{y}_2,x_2) \,.
\label{TypeIII}
\end{eqnarray}
As with Type II, the problem has been reduced to a sum of diagrams of
type I, each multiplied by an appropriate function of coordinates.
The contribution to (\ref{TypeIII}) at a given value of $k$ will have
the coordinate dependence
\begin{eqnarray}
(\vec{y}_{12}^2 + x_{12}^2)^{k-\Delta_2}\, { (2 x_2)^{(\Delta_1 -
\Delta_2+k) - k} \over (\vec{y}_{12}^2 + x_{12}^2)^{\Delta_1 -
\Delta_2 + k} } F(\xi) = { (2 x_2)^{\Delta_1 - \Delta_2} \over
(\vec{y}_{12}^2 + x_{12}^2)^{\Delta_1} } F(\xi) = {F(\xi)\over (2x_1
\xi)^{\Delta_1} (2x_2)^{\Delta_2}}\,,
\end{eqnarray}
which is indeed the proper form (\ref{ofourofour}) for a dCFT two-point 
function.

Many diagrams will also contain derivative couplings.  For example,
instead of (\ref{BulkInt}) consider the interaction
\begin{eqnarray}
\label{BulkIntDeriv}
S_{bulk} = {1 \over 2 \kappa_5^2} \int d^5z \sqrt{-g_5} \left(
\partial_\mu \phi_{\Delta_1} \partial^\mu \phi_{\Delta_2}
\phi_{\Delta_3} \right) \,,
\end{eqnarray}
along with the brane interaction (\ref{BraneInt}).
One finds
\begin{eqnarray}
{\cal M}_{\Delta_1 \Delta_2}(\vec{y}_1, x_1; \vec{y}_2, x_2) &=& \lambda  
\int {d^4w \over w_0^4} \, \tilde{A}(w;
\vec{y}_1, x_1; \vec{y}_2, x_2) \,, \\ \tilde{A}(w; \vec{y}_1,
x_1; \vec{y}_2, x_2) &\equiv& \int {d^5z \over
z_0^5} \, \partial_\mu K_{\Delta_1}(z; \vec{y}_1, x_1) \partial^\mu
K_{\Delta_2}(z; \vec{y}_2, x_2) \, G_{\Delta_3}(u)
\,, \nonumber
\end{eqnarray}
where both derivatives are with respect to the $z$-coordinate.  We can
process this by means of the identity (A.5) of \cite{dhoker} :
\begin{eqnarray}
\label{DerivIdentity}
\partial_\mu \tilde{K}_{\Delta_1}(z;1) \partial^\mu
\tilde{K}_{\Delta_2}(z;2) = \Delta_1 \Delta_2 \left[
\tilde{K}_{\Delta_1}(z;1) \tilde{K}_{\Delta_2}(z;2) - 2
(\vec{x}_{12}^2 + y_{12}^2) \tilde{K}_{\Delta_1+1 }(z;1)
\tilde{K}_{\Delta_2+1}(z;2) \right] \,.
\end{eqnarray}
This reduces the problem to two diagrams of the non-derivative type.
The former piece manifestly produces a result of the correct form
(\ref{ofourofour}), and it is easy to see that the latter piece does
as well.

\subsection{From correlators to anomalous dimensions}
\label{CorrAnomDimSec}

We have now seen that a wide class of two-point functions in a dCFT can
be expressed, in the limit of large 't Hooft coupling and to leading
order in the defect interactions, in terms of the functions
$I_{\Delta_1 \Delta_2}(\xi,x_1,x_2)$.  In subsection~\ref{TypeISec},
we saw that logarithms of $\xi$ occur in the large $\xi$ limit of
$I_{\Delta_1 \Delta_2}$ with
$\Delta_1+\Delta_2$ even. Like in the case of four-point functions in the
usual $AdS$/CFT computations \cite{dhoker}, these logarithms have an
interpretation in terms of corrections to scaling dimensions of
``intermediate states''.

As discussed in subsection~\ref{BOPESec}, a primary defect operator
$\co_n(\vec{y})$ of dimension $\Delta_n$ appearing in the BOPE of both
$\co_{\Delta_1}$ and $\co_{\Delta_2}$ will show up in the large-$\xi$
expansion of $\langle {\cal O}_{\Delta_1} \, {\cal O}_{\Delta_2}
\rangle$ as a term of order $1 / \xi^{\Delta_n}$. Thus, if the
dimension $\Delta_n$ is independent of $g_s M$ we should find a power
law behavior, while if $\Delta_n = \Delta_n^{(0)} + \Delta_n^{(1)} g_s M +
\ldots$ the expansion of the correlation function will behave as
\begin{eqnarray}
\label{loganomdim}
{1\over \xi^{\Delta_n}} = {1\over \xi^{\Delta_n^{(0)}}} (1 -
\Delta_n^{(1)} g_s M \log \xi + \ldots) \,.
\end{eqnarray}
In this way, subleading contributions to two-point functions can exhibit
logarithmic corrections at first order in $g_s M$. This is completely
analogous to the way that logarithms appear in four-point functions in
the standard computation of $AdS$/CFT correlation functions, where they
are related to anomalous dimensions of intermediate states appearing
in the OPE.

This argument suggests that such a subleading correction can only
appear if the leading contribution (at zeroth order in $g_s M$) is
non-zero, but actually this is not necessarily the case. If more than
one primary operator $\co_n$ of equal dimension $\Delta^{(0)}$ appears in the
BOPEs of $\co_{\Delta_1}$ and $\co_{\Delta_2}$, it is possible for the
leading contributions of the operators to cancel out precisely, and
then the second term in (\ref{loganomdim}) can actually be the leading
term in the two-point function (if the different primary 
operators have different $\Delta^{(1)}$'s).

In the case of four-point functions, anomalous dimensions arose for
intermediate composite operators of the form $[\co_1(x) \co_2(x)]$,
even when $\co_1$ and $\co_2$ themselves did not have anomalous
dimensions. This can occur because of the need to regularize the
operator product when the two operators approach each other, for
instance by a point-splitting regularization. The regularized
composite operator formed from two chiral operators need not be
chiral, and hence need not be protected from acquiring anomalous dimension.
In the dCFT case we similarly find anomalous dimensions for ``reduced
operators'' even when the ambient operators have no anomalous
dimensions.

For instance, consider an ambient operator $\co(\vec{y},x)$ of integer
dimension $\Delta$. The two-point function of this operator at zeroth
order in $g_s M$ is just the two-point function without the
defect. This has a simple BOPE interpretation in which the
intermediate states are just the coefficients in the Taylor expansion
of $\co$ around the defect, of dimensions $\Delta+n$ for $n=0,1,2,\ldots$.
Now, suppose that at first order in $g_s M$ we have a type I diagram
for the dual field.  This diagram gives a contribution proportional to
$I_{\Delta \Delta}$ to the two-point function, and the leading
large-$\xi$ contribution of this diagram to the correlation function
behaves as $\lambda \log \xi / \xi^{\Delta} x_1^{\Delta} x_2^{\Delta}$
with $\lambda \propto g_s M$. Using (\ref{loganomdim}) we interpret
such a contribution as resulting from the fact that the reduced
operator $[\co](\vec{y}) \equiv \lim_{x\to 0} \co(\vec{y},x) $, which
had dimension $\Delta$ at leading order in $g_s M$, now acquires an
anomalous dimension proportional to $\lambda$.  Again, it is natural
to interpret the fact that the reduced operator $[\co](\vec{y})$
acquires an anomalous dimension even when the ambient operator
$\co(\vec{y},x)$ does not as related to the fact that the limit
$\lim_{x\to 0} \co(\vec{y},x)$ is non-trivial once interactions with
the defect are introduced, and a regularization method such as
point-splitting is required to define the reduced operator, which can
lead to an anomalous dimension for these operators.

Similarly, the higher powers of $1/\xi$ in the expansion of two-point
functions can be related to anomalous dimensions of the higher reduced
operators $[\partial_x^n \co](\vec{y}) \equiv \lim_{x\to 0} \del_x^n
\co(\vec{y},x)$. In general, operators involving the defect fields
will also appear in the BOPE expansion of the two-point function, but
they cannot appear in the two-point function at first order in $g_s M$
since their BOPE coefficients are themselves of order $g_s M$. Thus,
at leading order in $g_s M$ the logarithmic terms in the two-point
functions just teach us about the anomalous dimensions of the reduced
operators. As discussed above, these are relevant for identifying
locally localized fields.

We can also discuss the small-$\xi$ limit of the correlation functions
we find. As described in section \ref{bdod}, this is related to
one-point functions of intermediate operators appearing in the ambient
OPE of $\co_{\Delta_1}$ and $\co_{\Delta_2}$. This interpretation is
particularly clear for the case of type III diagrams. These diagrams
involve a bulk coupling of some field $\phi_{\Delta_3}$ to
$\phi_{\Delta_1} \phi_{\Delta_2}$, which is related to the OPE
coefficient of the corresponding operator $\co_{\Delta_3}$ in the OPE
of $\co_{\Delta_1}$ with $\co_{\Delta_2}$, and they involve a
one-point function of $\phi_{\Delta_3}$ on the brane, which is
directly related to the one-point function of $\co_{\Delta_3}$. It is
easy to check that the coefficients involved in these relations all
match, so the small-$\xi$ limit of these diagrams agrees with our
expectations. The interpretation of the small-$\xi$ behavior of type I
and II diagrams is less straightforward; the leading operator visible
in the OPE expansion of diagrams of these types has dimension
$2\Delta$, and can be identified with $[\co_\Delta \co_\Delta]$.

\section*{Acknowledgments}
 
We would like to thank Allan Adams, Neil Constable, Steve Giddings,
Chris Herzog, Hirosi Ooguri, Joe Polchinski, Lisa Randall and Johannes
Walcher for useful discussions. We are grateful to the Aspen Center
for Physics for hospitality during various stages of this project in
the summers of 2001 and 2002. OA would also like to thank Harvard
University, the University of British Columbia, Stanford University
and SLAC for hospitality. The work of OA was supported in part by the
Israel-U.S. Binational Science Foundation, by the ISF Centers of
Excellence program, by the European network HPRN-CT-2000-00122, and by
Minerva.  OA is the incumbent of the Joseph and Celia Reskin career
development chair.  OD would like to thank Harvard University and SLAC
where parts of this work were done.  The research of OD was supported
by the National Science Foundation under grant PHY99-07949.  The work
of DZF was supported by the National Science Foundation under grant
PHY00-96515.  The work of AK was partially supported by the DOE under
contract DE-FGO3-96-ER40956.

\begin{appendix}
\section{Field theory conventions}

Here we present the propagators necessary for evaluating the
correlation functions in section \ref{anom_dim}. The toy dQFT consists
of the $d=4$ fermions $\lambda_{im}$ and the defect fields $\Psi^i$,
$q^m$, ($i,m=1,2$).  We are not careful about whether the indices $m$,
$i$ are raised or lowered, but move them around for notational
convenience. We use the spinor conventions of \cite{dfo}.

The $\Psi^i$ and $q^m$ are complex.  The $\lambda_{im}$ are composed
of Majorana fields $\lambda$, $\chi^A$, $A = 1,2,3$:
\begin{eqnarray}
\lambda_{im} \equiv \lambda \delta_{im} - i \chi^A \sigma^A_{im} \,,
\quad \quad  \bar\lambda_{mi} \equiv \bar\lambda \delta_{mi} + i \bar\chi^A
\sigma^A_{mi} \,,
\end{eqnarray}
and the kinetic terms for the $\lambda$, $\chi^A$ are canonical:
\begin{eqnarray}
\label{LambdaKin}
-\tf{i}{4} \bar\lambda^{mi} \gamma^\mu \partial_\mu \lambda^{im} =
-\tf{i}{2} \bar\lambda \gamma^\mu \partial_\mu \lambda -\tf{i}{2}
\bar\chi^A \gamma^\mu \partial_\mu \chi^A \,.
\end{eqnarray}
The action (\ref{Action}) has three continuous symmetries, the defect
$U(1)_B$, under which $q$ and $\Psi$ both have charge 1, and $SU(2)_V
\times SU(2)_H$ (acting on the indices $i$ and $m$, respectively),
under which $\Psi \rightarrow g_V \Psi$, $q \rightarrow g_H q$, and
$\lambda \rightarrow g_V \lambda g_H^\dagger$, with $T^A = \sigma^A/2$
the generators of $SU(2)$, leading to the currents (\ref{currents}).

We now consider the propagators, rotated to Euclidean space.  For
$\lambda_{im}$, one processes the $SU(2)_V \times SU(2)_H$ indices
according to
\begin{eqnarray}
\nonumber
\langle \lambda_{im}(z_1) \, \bar\lambda_{nj}(z_2) \rangle &=&  2 \langle 
\lambda(z_1) \, \bar\lambda(z_2) \rangle \delta_{ij} \delta_{mn} \,, \\
\label{FlavorProp}
\langle \lambda_{im}(z_1) \, \lambda_{jn}(z_2) \rangle &=&  2 \langle 
\lambda(z_1) \, \lambda(z_2) \rangle \epsilon_{ij} \epsilon_{mn} \,, \\
\langle \bar\lambda_{mi}(z_1) \, \bar\lambda_{nj}(z_2) \rangle &=&  2 \langle 
\bar\lambda(z_1) \, \bar\lambda(z_2) \rangle \epsilon_{ij} \epsilon_{mn} \,,
\nonumber
\end{eqnarray}
where we suppressed spinor indices; the propagators on the right-hand
side those of an ordinary Majorana particle, evaluated below.  In
evaluating the above we used the Pauli matrix identity:
\begin{eqnarray}
\sigma^A_{ij} \sigma^A_{kl} = \delta_{ij} \delta_{kl} - 2
\epsilon_{ik} \epsilon_{jl} = 2 \delta_{il} \delta_{jk} - 
\delta_{ij} \delta_{kl} \,.
\end{eqnarray}
The massless 4D and 3D scalar propagators
are:\footnote{Of course there is no field $X$ in this model, but it is
nonetheless useful to define the 4D scalar propagator.}
\begin{eqnarray}
\langle X(z_1) \, \bar{X}(z_2) \rangle \equiv \Delta_{12} =  
{1 \over 4 \pi^2} {
1 \over (z_{12})^2} \,, \quad \quad
\langle q(y_1) \, \bar{q}(y_2) \rangle \equiv \delta_{12} =  
{1 \over 4 \pi} {1 
\over |y_{12}|} \,.
\end{eqnarray}
The massless 4D Fermi propagator is:
\begin{eqnarray}
\langle \lambda(z_1) \, \bar{\lambda}(z_2) \rangle \equiv
S_{12} = \gamma^\mu\partial_\mu \Delta_{12}  \,.
\end{eqnarray}
Since $\lambda$ is Majorana, we also have nonvanishing $\langle
\lambda_{\alpha a}(z_1) \, \lambda_{\beta b} (z_2) \rangle$ and
$\langle \bar\lambda_{\alpha a}(z_1) \, \bar\lambda_{\beta b}(z_2)
\rangle$, determined by $\bar\lambda = \lambda^T \gamma^0$.
The massless 3D Fermi propagator is:
\begin{eqnarray}
\langle \Psi (y_1) \, \bar{\Psi} (y_2) \rangle \equiv 
s_{12} = \rho^k  \partial_k \, \delta_{12} \,,
\end{eqnarray}
which can be written as in a 4D notation as
\begin{eqnarray}
\label{PsiProp}
\langle \Psi^i(z_1) \bar\Psi^j(z_2) \rangle = \delta^{ij}
(\gamma^k P_+)  \,\partial_k \delta_{12} \equiv \delta^{ij}
\hat{s}_{12} \,,
\end{eqnarray}
where we define the projection matrices $P_\pm$ 
\begin{eqnarray}
\label{ProjectionMatrices}
P_\pm \equiv \tf12 \left( 1 \pm \gamma^5 \gamma^x \right) \,,
\end{eqnarray}
where $\gamma^5 \equiv -i \gamma^0 \gamma^1 \gamma^2 \gamma^x$ is the
chirality matrix. The projection matrices (\ref{ProjectionMatrices}) obey
\begin{eqnarray}
[P_\pm , \gamma^k ] = 0 \,, \quad  \gamma^x P_\pm = P_\mp \gamma^x \,, \quad 
P_+^2 = P_-^2 = 1 \,, \quad P_+ P_- = 0 \,.
\end{eqnarray}

\end{appendix}

\bibliography{adfk}

\begingroup\raggedright\begin{thebibliography}{10}

\bibitem{Rubakov}
V.~A. Rubakov and M.~E. Shaposhnikov, ``Do we live inside a domain wall?,''
  {\em Phys. Lett.} {\bf B125} (1983) 136--138.

\bibitem{rs2}
L.~Randall and R.~Sundrum, ``An alternative to compactification,'' {\em Phys.
  Rev. Lett.} {\bf 83} (1999) 4690--4693,
  \href{http://xxx.lanl.gov/abs/hep-th/9906064}{{\tt hep-th/9906064}}.

\bibitem{kaloper}
N.~Kaloper, ``Bent domain walls as braneworlds,'' {\em Phys. Rev.} {\bf D60}
  (1999) 123506, \href{http://xxx.lanl.gov/abs/hep-th/9905210}{{\tt
  hep-th/9905210}}.

\bibitem{kimkim}
H.~B. Kim and H.~D. Kim, ``Inflation and gauge hierarchy in Randall-Sundrum
  compactification,'' {\em Phys. Rev.} {\bf D61} (2000) 064003,
  \href{http://xxx.lanl.gov/abs/hep-th/9909053}{{\tt hep-th/9909053}}.

\bibitem{nihei}
T.~Nihei, ``Inflation in the five-dimensional universe with an orbifold extra
  dimension,'' {\em Phys. Lett.} {\bf B465} (1999) 81--85,
  \href{http://xxx.lanl.gov/abs/hep-ph/9905487}{{\tt hep-ph/9905487}}.

\bibitem{dfgk}
O.~DeWolfe, D.~Z. Freedman, S.~S. Gubser, and A.~Karch, ``Modeling the fifth
  dimension with scalars and gravity,'' {\em Phys. Rev.} {\bf D62} (2000)
  046008, \href{http://xxx.lanl.gov/abs/hep-th/9909134}{{\tt hep-th/9909134}}.

\bibitem{kr}
A.~Karch and L.~Randall, ``Locally localized gravity,'' {\em JHEP} {\bf 05}
  (2001) 008, \href{http://xxx.lanl.gov/abs/hep-th/0011156}{{\tt
  hep-th/0011156}}.

\bibitem{juan}
J.~Maldacena, ``The large $N$ limit of superconformal field theories and
  supergravity,'' {\em Adv. Theor. Math. Phys.} {\bf 2} (1998) 231--252,
  \href{http://xxx.lanl.gov/abs/hep-th/9711200}{{\tt hep-th/9711200}}.

\bibitem{magoo}
O.~Aharony, S.~S. Gubser, J.~M. Maldacena, H.~Ooguri, and Y.~Oz, ``Large N
  field theories, string theory and gravity,'' {\em Phys. Rept.} {\bf 323}
  (2000) 183--386, \href{http://xxx.lanl.gov/abs/hep-th/9905111}{{\tt
  hep-th/9905111}}.

\bibitem{kr2}
A.~Karch and L.~Randall, ``Open and closed string interpretation of SUSY CFT's
  on branes with boundaries,'' {\em JHEP} {\bf 06} (2001) 063,
  \href{http://xxx.lanl.gov/abs/hep-th/0105132}{{\tt hep-th/0105132}}.

\bibitem{dfo}
O.~DeWolfe, D.~Z. Freedman, and H.~Ooguri, ``Holography and defect conformal
  field theories,'' {\em Phys. Rev.} {\bf D66} (2002) 025009,
  \href{http://xxx.lanl.gov/abs/hep-th/0111135}{{\tt hep-th/0111135}}.

\bibitem{erdmenger1}
J.~Erdmenger, Z.~Guralnik, and I.~Kirsch, ``Four-dimensional superconformal
  theories with interacting boundaries or defects,'' {\em Phys. Rev.} {\bf D66}
  (2002) 025020, \href{http://xxx.lanl.gov/abs/hep-th/0203020}{{\tt
  hep-th/0203020}}.

\bibitem{sav2}
A.~Kapustin and S.~Sethi, ``The Higgs branch of impurity theories,'' {\em Adv.
  Theor. Math. Phys.} {\bf 2} (1998) 571--591,
  \href{http://xxx.lanl.gov/abs/hep-th/9804027}{{\tt hep-th/9804027}}.

\bibitem{bp}
C.~Bachas and M.~Petropoulos, ``Anti-de-Sitter D-branes,'' {\em JHEP} {\bf 02}
  (2001) 025, \href{http://xxx.lanl.gov/abs/hep-th/0012234}{{\tt
  hep-th/0012234}}.

\bibitem{bbdo}
C.~Bachas, J.~de~Boer, R.~Dijkgraaf, and H.~Ooguri, ``Permeable conformal walls
  and holography,'' {\em JHEP} {\bf 06} (2002) 027,
  \href{http://xxx.lanl.gov/abs/hep-th/0111210}{{\tt hep-th/0111210}}.

\bibitem{Quella}
T.~Quella and V.~Schomerus, ``Symmetry breaking boundary states and defect
  lines,'' {\em JHEP} {\bf 06} (2002) 028,
  \href{http://xxx.lanl.gov/abs/hep-th/0203161}{{\tt hep-th/0203161}}.

\bibitem{Lee}
P.~Lee and J.-w. Park, ``Open strings in PP-wave background from defect
  conformal field theory,'' {\em Phys. Rev.} {\bf D67} (2003) 026002,
  \href{http://xxx.lanl.gov/abs/hep-th/0203257}{{\tt hep-th/0203257}}.

\bibitem{Skenderis}
K.~Skenderis and M.~Taylor, ``Branes in AdS and pp-wave spacetimes,'' {\em
  JHEP} {\bf 06} (2002) 025, \href{http://xxx.lanl.gov/abs/hep-th/0204054}{{\tt
  hep-th/0204054}}.

\bibitem{Mateos}
D.~Mateos, S.~Ng, and P.~K. Townsend, ``Supersymmetric defect expansion in CFT
  from AdS supertubes,'' {\em JHEP} {\bf 07} (2002) 048,
  \href{http://xxx.lanl.gov/abs/hep-th/0207136}{{\tt hep-th/0207136}}.

\bibitem{Yamaguchi}
S.~Yamaguchi, ``Holographic RG flow on the defect and g-theorem,'' {\em JHEP}
  {\bf 10} (2002) 002, \href{http://xxx.lanl.gov/abs/hep-th/0207171}{{\tt
  hep-th/0207171}}.

\bibitem{erdmenger2}
N.~R. Constable, J.~Erdmenger, Z.~Guralnik, and I.~Kirsch, ``Intersecting
  D3-branes and holography,''
  \href{http://xxx.lanl.gov/abs/hep-th/0211222}{{\tt hep-th/0211222}}.

\bibitem{erdmenger3}
N.~R. Constable, J.~Erdmenger, Z.~Guralnik, and I.~Kirsch, ``(De)constructing
  intersecting M5-branes,'' \href{http://xxx.lanl.gov/abs/hep-th/0212136}{{\tt
  hep-th/0212136}}.

\bibitem{fayya}
A.~Fayyazuddin, ``Supersymmetric webs of D3/D5-branes in supergravity,''
  \href{http://xxx.lanl.gov/abs/hep-th/0207129}{{\tt hep-th/0207129}}.

\bibitem{HW}
A.~Hanany and E.~Witten, ``Type IIB superstrings, BPS monopoles, and
  three-dimensional gauge dynamics,'' {\em Nucl. Phys.} {\bf B492} (1997)
  152--190, \href{http://xxx.lanl.gov/abs/hep-th/9611230}{{\tt
  hep-th/9611230}}.

\bibitem{witten}
E.~Witten, ``Anti-de Sitter space and holography,'' {\em Adv. Theor. Math.
  Phys.} {\bf 2} (1998) 253--291,
  \href{http://xxx.lanl.gov/abs/hep-th/9802150}{{\tt hep-th/9802150}}.

\bibitem{wrt}
E.~D'Hoker, D.~Z. Freedman, and L.~Rastelli, ``AdS/CFT 4-point functions: How
  to succeed at z-integrals without really trying,'' {\em Nucl. Phys.} {\bf
  B562} (1999) 395--411, \href{http://xxx.lanl.gov/abs/hep-th/9905049}{{\tt
  hep-th/9905049}}.

\bibitem{mcao}
D.~M. McAvity and H.~Osborn, ``Conformal field theories near a boundary in
  general dimensions,'' {\em Nucl. Phys.} {\bf B455} (1995) 522--576,
  \href{http://xxx.lanl.gov/abs/cond-mat/9505127}{{\tt cond-mat/9505127}}.

\bibitem{BKL}
V.~Balasubramanian, P.~Kraus, and A.~E. Lawrence, ``Bulk vs. boundary dynamics
  in anti-de Sitter spacetime,'' {\em Phys. Rev.} {\bf D59} (1999) 046003,
  \href{http://xxx.lanl.gov/abs/hep-th/9805171}{{\tt hep-th/9805171}}.

\bibitem{por1}
M.~Porrati, ``Mass and gauge invariance. IV: Holography for the Karch- Randall
  model,'' {\em Phys. Rev.} {\bf D65} (2002) 044015,
  \href{http://xxx.lanl.gov/abs/hep-th/0109017}{{\tt hep-th/0109017}}.

\bibitem{matthew}
M.~D. Schwartz, ``The emergence of localized gravity,'' {\em Phys. Lett.} {\bf
  B502} (2001) 223--228, \href{http://xxx.lanl.gov/abs/hep-th/0011177}{{\tt
  hep-th/0011177}}.

\bibitem{miemiec}
A.~Miemiec, ``A power law for the lowest eigenvalue in localized massive
  gravity,'' {\em Fortsch. Phys.} {\bf 49} (2001) 747--755,
  \href{http://xxx.lanl.gov/abs/hep-th/0011160}{{\tt hep-th/0011160}}.

\bibitem{diffreg}
D.~Z. Freedman, K.~Johnson, and J.~I. Latorre, ``Differential regularization
  and renormalization: A New method of calculation in quantum field theory,''
  {\em Nucl. Phys.} {\bf B371} (1992) 353--414.

\bibitem{dhoker}
E.~D'Hoker, D.~Z. Freedman, S.~D. Mathur, A.~Matusis, and L.~Rastelli,
  ``Graviton exchange and complete 4-point functions in the AdS/CFT
  correspondence,'' {\em Nucl. Phys.} {\bf B562} (1999) 353--394,
  \href{http://xxx.lanl.gov/abs/hep-th/9903196}{{\tt hep-th/9903196}}.

\end{thebibliography}\endgroup
\bibliographystyle{ssg}

\end{document}